\documentclass[floats,floatfix,amssymb,prd,twocolumn,nofootinbib,superscriptaddress]{revtex4-1}
\usepackage{amssymb,amsmath,verbatim,mathtools,needspace,enumitem,etoolbox,graphicx,physics,microtype,ulem}
\usepackage{natbib,tabularx}
\usepackage{xspace}
\usepackage[dvipsnames, usenames]{xcolor}
\usepackage{booktabs}%for tables
\newcolumntype{Y}{>{\hsize=0.9\hsize\centering\arraybackslash}X}
\newcolumntype{W}{>{\hsize=2.8\hsize\centering\arraybackslash}X}
\newcolumntype{N}{>{\hsize=0.7\hsize\centering\arraybackslash}X}
\newcolumntype{b}{>{\hsize=1.6\hsize\centering\arraybackslash}X}
\newcolumntype{Z}{>{\hsize=0.6\hsize\centering\arraybackslash}X} %have to add up to totally number of cols

\definecolor{linkcolor}{rgb}{0.0,0.3,0.5}
\definecolor{urlcolor}{rgb}{0.27,0.55,0.}
\definecolor{funcolor}{rgb}{0.65, 0.16, 0.16}
\usepackage[unicode, colorlinks=true, linkcolor=linkcolor, citecolor=linkcolor, filecolor=linkcolor,urlcolor=linkcolor, pdfusetitle]{hyperref}
\hypersetup{colorlinks=true}
\usepackage[all]{hypcap}
\usepackage{aas_macros}
\usepackage{subcaption}

\newcommand{\chieff}{\ensuremath{\chi_{\mathrm{eff}}}\xspace}
\newcommand{\chip}{\ensuremath{\chi_{\mathrm{p}}}\xspace}
\newcommand{\mcsource}{\ensuremath{\mathcal{M}^{\mathrm{source}}}\xspace}

\newcommand{\degg}{\ensuremath{\mathrm{deg}^2}\xspace}

\newcommand{\beqn}{\begin{eqnarray}}
\newcommand{\enqn}{\end{eqnarray}}
\newcommand{\beq}{\begin{equation}}
\newcommand{\eeq}{\end{equation}}

\usepackage{gensymb} %
\usepackage{amsbsy} %
\usepackage{cleveref}

\begin{document}
\title{Characterization of low-significance gravitational-wave compact binary sources}

\author{Yiwen Huang}
\email{ywh@mit.edu}
\affiliation{LIGO, Massachusetts Institute of Technology, Cambridge, Massachusetts 02139, USA}
\affiliation{Department of Physics and Kavli Institute for Astrophysics and Space Research, Massachusetts Institute of Technology, Cambridge, Massachusetts 02139, USA}

\author{Hannah Middleton}
\affiliation{Institute of Gravitational Wave Astronomy and School of Physics and Astronomy, University of Birmingham, Birmingham, B15 2TT, United Kingdom}
\affiliation{OzGrav-Melbourne, University of Melbourne, School of Physics, Parkville, Victoria 3010, Australia}

\author{Ken K.~Y. Ng}%
\affiliation{LIGO, Massachusetts Institute of Technology, Cambridge, Massachusetts 02139, USA}
\affiliation{Department of Physics and Kavli Institute for Astrophysics and Space Research, Massachusetts Institute of Technology, Cambridge, Massachusetts 02139, USA}

\author{Salvatore Vitale}
\email{salvatore.vitale@ligo.org}
\affiliation{LIGO, Massachusetts Institute of Technology, Cambridge, Massachusetts 02139, USA}
\affiliation{Department of Physics and Kavli Institute for Astrophysics and Space Research, Massachusetts Institute of Technology, Cambridge, Massachusetts 02139, USA}

\author{John Veitch}
\affiliation{Institute for Gravitational Research, School of Physics and Astronomy, University of Glasgow, Glasgow, G12 8QQ, United Kingdom}

\date{\today}

\begin{abstract}

Advanced LIGO and Virgo have so far detected gravitational waves from 10 binary black hole mergers (BBH) and 1 binary neutron star merger (BNS).
In the future, we expect the detection of many more marginal sources, since compact binary coalescences detectable by advanced ground-based instruments are roughly distributed uniformly in comoving volume. In this paper we simulate weak signals from compact binary coalescences of various morphologies and optimal network signal-to-noise ratios (henceforth SNRs), and analyze if and to which extent their parameters can be measured by advanced LIGO and Virgo in their third observing run.
We show that subthreshold binary neutron stars, with SNRs below 12 (10) yield uncertainties in their sky position larger than 400 (700)~\degg (90\% credible interval). The luminosity distance, which could be used to measure the Hubble constant with standard sirens, has relative uncertainties larger than 40\% for BNSs and neutron star black hole mergers. For sources with SNRs below 8, it is not uncommon that the extrinsic parameters, sky position and distance, cannot be measured. 
Next, we look at the intrinsic parameters, masses and spins. We show that the detector-frame chirp mass can sometimes be measured with uncertainties below $1\%$ even for sources at SNRs of 6, although multimodality is not uncommon and can significantly broaden the posteriors. 
The effective inspiral spin is best measured for neutron star black hole mergers, for which the uncertainties can be as low as $\sim0.08$ ($\sim 0.2$) at SNR 12 (8). The uncertainty is higher for systems with comparable component masses or lack of spin precession. 
\end{abstract}

\maketitle

\section{Introduction}

Interferometric ground-based gravitational-wave (GW) detectors, such as Advanced LIGO~\cite{Harry:2010zz, TheLIGOScientific:2014jea} and Advanced Virgo~\cite{TheVirgo:2014hva}
are most sensitive to GWs in the frequency band of $[20-2,000]$~Hz~\cite{ligosen}. 
The merger of compact objects such as neutron stars and stellar-mass black holes in binaries can produce GW signals in this frequency band, and will be the most commonly detected sources by ground-based detectors. 
At the time of writing this paper, LIGO and Virgo have published the detections of 10 BBHs~\cite{O1-BBH, GW150914-DETECTION, GW151226-DETECTION, 2017PhRvL.118v1101A, 2017ApJ...851L..35A, 2017PhRvL.119n1101A,LIGOScientific:2018jsj} and one BNS, GW170817~\cite{2017PhRvL.119p1101A,Abbott:2018wiz}.

These first few detections have already made clear how characterization of compact binary coalescence (CBC) sources will broaden our understanding of the key properties of black holes and neutron stars. As more detections are made, they will make it possible to constrain the role of stellar wind, rotation, and metallicity in the progenitor stars~\cite{GW150914-ASTRO}, to measure the merger rate~\cite{MergerRate150914}, the spin and mass distribution~\cite{Vitale:2015tea,2017arXiv170907896F,2017Natur.548..426F,2017MNRAS.471.2801S,2016arXiv160808223M,PhysRevLett.112.251101,Vitale:2016avz,2009CQGra..26t4010V} and to perform  stringent tests of general relativity in its strong-field regime~\cite{GW150914-TESTOFGR,2016PhRvL.117f1102L,2016PhRvD..94b1101G,2016arXiv160308955Y,2014PhRvD..89h2001A,2012PhRvD..85h2003L}.

Some of the expected sources of GWs are also luminous in the electromagnetic (EM) band, which allow for the possibility of multimessenger observations. This has been spectacularly shown with the joint detection of GW and EM signals from the BNS source GW170817~\cite{2017ApJ...848L..12A}. The host of GW170817 was identified, together with radiation in the whole EM spectrum, from radio to $\gamma$-rays~\cite{2017Sci...358.1556C,2017ApJ...848L..12A, GW170817-GRB,2017ApJ...848L..16S,2017Sci...358.1579H}. The science output of this discovery is too rich to be fully described here. We thus only mention a few highlights. GW170817 was used to set constraints on the equation of state of neutron stars~\cite{Abbott:2018exr}, search for evidence of p-g modes~\cite{Weinberg:2018icl} and put bounds on the component neutron star masses and spins~\cite{Abbott:2018wiz}. The EM data confirmed the connection between short gamma-ray bursts and BNS sources, lead to the observation of the kilonova, and yielded insights on the details of the EM emission~\cite{2017Sci...358.1570D,2018ApJ...856..101M,2017ApJ...851L..21V,2017Sci...358.1565E}. Information from both the GW and the EM sides was used to measure the Hubble constant in a way that is independent of the cosmic distance ladder~\cite{HubbleConstant}. 
As more BNSs are detected in the next years, we will be able to gain a more solid understanding of the properties of compact binaries, their progenitors, and the electromagnetic radiation they emit.

Given the current detections, it is possible to estimate the local merger rates of binary neutron stars~\cite{2017PhRvL.119p1101A} and binary black holes~\cite{MergerRate150914,O1-BBH,2017PhRvL.118v1101A}. Neutron star black hole mergers (NSBHs) are also promising sources, but have not been detected yet, hence there are only predictions for their merger rate~\cite{MergerRateNSBH}. The measured or predicted merger rates together with the projected improvements in the sensitivity of LIGO and Virgo~\cite{2016LRR....19....1A} allow one to estimate the number of detections per year above some pre-defined matched-filter signal-to-noise ratio (SNR) or false alarm rate threshold~\cite{2016LRR....19....1A}.

As long as the distance distribution of GW sources can be considered uniform in volume (which is a good approximation for sources close enough that cosmological effects can be neglected), the distribution of events' luminosity distances  $d_L$ is proportional to $d_L^2$. Given that the SNR $\rho$ goes like $1/d_L$, the resulting SNR distribution goes like $\rho^{-4}$~\cite{2011CQGra..28l5023S}.
Even a smaller decrease in the threshold matched-filter SNR should thus correspond to a large increase in the number of detected events. 
In practice, other factors limit the benefit of lowering the detection threshold.
The noise from GW detectors is not perfectly gaussian. Non-gaussian noise artifacts (often referred to as \emph{glitches}) limit the sensitivity of GW searches~\cite{GW150914-DETCHAR}.  To assess the significance of a candidate event, its SNR (or other detection statistics) is compared with the distribution of SNRs from the background, which is usually estimated either by time-sliding the data of different instruments~\cite{2016CQGra..33u5004U} or by constructing the network SNR distribution assuming noise is independent in each detector~\cite{2013PhRvD..88b4025C}.
Unfortunately, while the distribution of SNRs from CBCs goes like $\rho^{-4}$, the distribution of SNRs from glitches increases much faster as the SNR decreases~\cite{PhysRevD.85.082002,2017ApJ...849..118N}

This can limit the benefits of lowering the threshold while searching for CBC signals and following them up in the EM band. For example, Ref.~\cite{Lynch:2018yom} has shown that lowering the threshold false alarm rate to 1 per month (week) would result in only 39\% (13\%) of the BNS candidate to be of astrophysical origin. Observers who decide to follow up marginal events would thus have to deal with a large number of false positives. 

In this paper we look at a complementary aspect of marginal events, namely the fact that their characterization can be challenging or inconclusive. Virtually all of the GW literature focusing on characterization of GWs from compact binaries has considered clear detections, generating simulated GW signals with optimal SNRs above some threshold (often $\sim 12$). 
In this paper we reverse that approach, and only consider sources \emph{weaker} than what could be considered a clear detection. 

Since the Fisher matrix approximation fails at small SNRs~\cite{2008PhRvD..77d2001V}, we perform full numerical simulations. We simulate BNS, BBH, and NSBH sources and add them to real interferometric noise of LIGO and Virgo, recolored to have the spectral behavior expected in the next observing run, starting in early 2019~\cite{2016CQGra..33u5004U}.
We analyze the sources at different optimal network SNRs~\footnote{The square of the optimal network SNR is defined as the sum of the squares of the single-instrument optimal SNR, for all instruments taking data. The optimal SNR is defined in Eq.~\ref{Eq.SNR}.}, from 6 to 12, and show exactly how the quality of the parameter estimation process degrades as the optimal network SNR decreases. We find even at optimal network SNR of 12, BNS sources cannot be localized to areas smaller than $\sim 400$~\degg (90\% credible interval). Meanwhile, the luminosity distance is always measured with relative uncertainties larger than $40\%$. It is not uncommon that the sky position and luminosity distance simply cannot be measured.
This reveals the challenges of finding an EM counterpart to a weak GW source, and to make a compelling case that the association is real.
The chirp mass, which is usually the best measured parameter, shows signs of multimodality, especially for heavy systems. At optimal network SNR 7 or below, for the majority of the sources we consider, we obtain posterior distributions for the chirp mass which is multimodal, or with very large tails. The uncertainty in the mass ratio is large for heavy systems, whereas it can be as small as $0.07$ for NSBH at optimal network SNR 12.
Finally, we show that, especially for NSBH, the effective spin~\cite{PhysRevD.64.124013,PhysRevD.78.044021,PhysRevD.82.064016,2011PhRvL.106x1101A} can be measured at very low optimal network SNRs.

The rest of the paper is structured as follows: in Sec. ~\ref{Method}, we report the setup of the simulations we performed, including noise and signal generation, and of the analysis. In Sec. ~\ref{Results}, we describe the main results for a set of simulated weak sources. The conclusions are presented in Sec. ~\ref{Sec.Conclusions}.
\section{Method}
\label{Method}
\subsection{Noise and signal generation}\label{signal}

The data in each interferometer in presence of a GW signal can be written as $\boldsymbol {d}$ :
\begin{equation}
\boldsymbol {d} = \boldsymbol {h} + \boldsymbol {n}
\end{equation}
where $\boldsymbol {h}$ is signal and $\boldsymbol {n}$ is noise. 

The necessary ingredients to simulate an end-to-end analysis of a GW signal are thus the generation of a synthetic GW signal, and a stretch of data that the signal can be added to. In this section we describe our approach to generate these quantities, starting from the noise.

We work with a network made of the two advanced LIGO instruments and the advanced Virgo detector.
To make this study immediately relevant to the next observing run (O3, starting in early 2019~\cite{ligosen}), we work with noise streams that have the projected O3 sensitivity for each instrument~\footnote{However, the results we present will be relevant for any situations in which the two LIGO detectors accumulate most of the network SNR, which will likely be the case for the next few years~\cite{2016LRR....19....1A}.}. The power spectral density (PSD), $S_n(f)$, of a stretch of data is defined as the average of the noise autocorrelation over the duration of the data segment~\cite{SathyaSchutzLRR}. It is trivial to produce Gaussian noise colored to have any specific PSD, and this is the approach followed in a significant fraction of the gravitational-wave literature. However, when dealing with marginal events, using real noise seems important, as small noise artifacts, or fluctuations, might have significant impact on weak signals.

We use public real data from the first observing run (O1)~\cite{GWOSC} and recolor it to have the expected O3 spectral behavior. More specifically, we select 5 GPS times in O1 such that the data around them does not contain any (known) astrophysical events nor significant instrumental artifact that would have resulted in vetoing of that data (Ref.~\cite{GWOSC} provides list of data segments which are considered clean by the LIGO and Virgo collaboration). The O1 times we used for our analysis are listed in Table~\ref{Table.GPS}.
We download the corresponding 5 data files for each of the LIGO instruments from the Ref.~\cite{GWOSC}  and use a routine of the \textit{gstlal} algorithm~\cite{Cannon2010,2011PhRvD..84h4003C,2013PhRvD..88b4025C,2015arXiv150404632C,2017PhRvD..95d2001M} to apply the expected O3 power spectrum (specifically the lower bounds of the ``Late'' curve (2018-19) for Advanced LIGO and of the ``Mid'' curve (2018-19) for Advanced Virgo in~\cite{ligosen}.).

Since Virgo was not running together with the advanced LIGO in O1, no Virgo O1 data is available for those 5 GPS times. 
We simulate Virgo data as follows. For each GPS time, we pick a Hanford frame corresponding to one of the \emph{other} four GPS times, shift the time stamp to coincide with the desired GPS time, and recolor it to the projected O3 Virgo PSD.

We stress that  while recoloring O1 data to a target sensitivity (O3, in this case) gives a way to make prediction about the performances of future science runs, it also has limitations. In particular, it will not capture new types of instrumental artifacts that might arise as the instruments get more sensitive. On the other hand, recoloring archival data maintains old instrumental artifacts that might very well be solved in the future. However, as our signals do not overlap with major instrumental artifacts, and given the lack of alternatives, we proceed with recolored data, and use Hanford data as time-shifted mock Virgo data.

\begin{figure}[htb]
  \centering
    \includegraphics[width=0.45\textwidth]{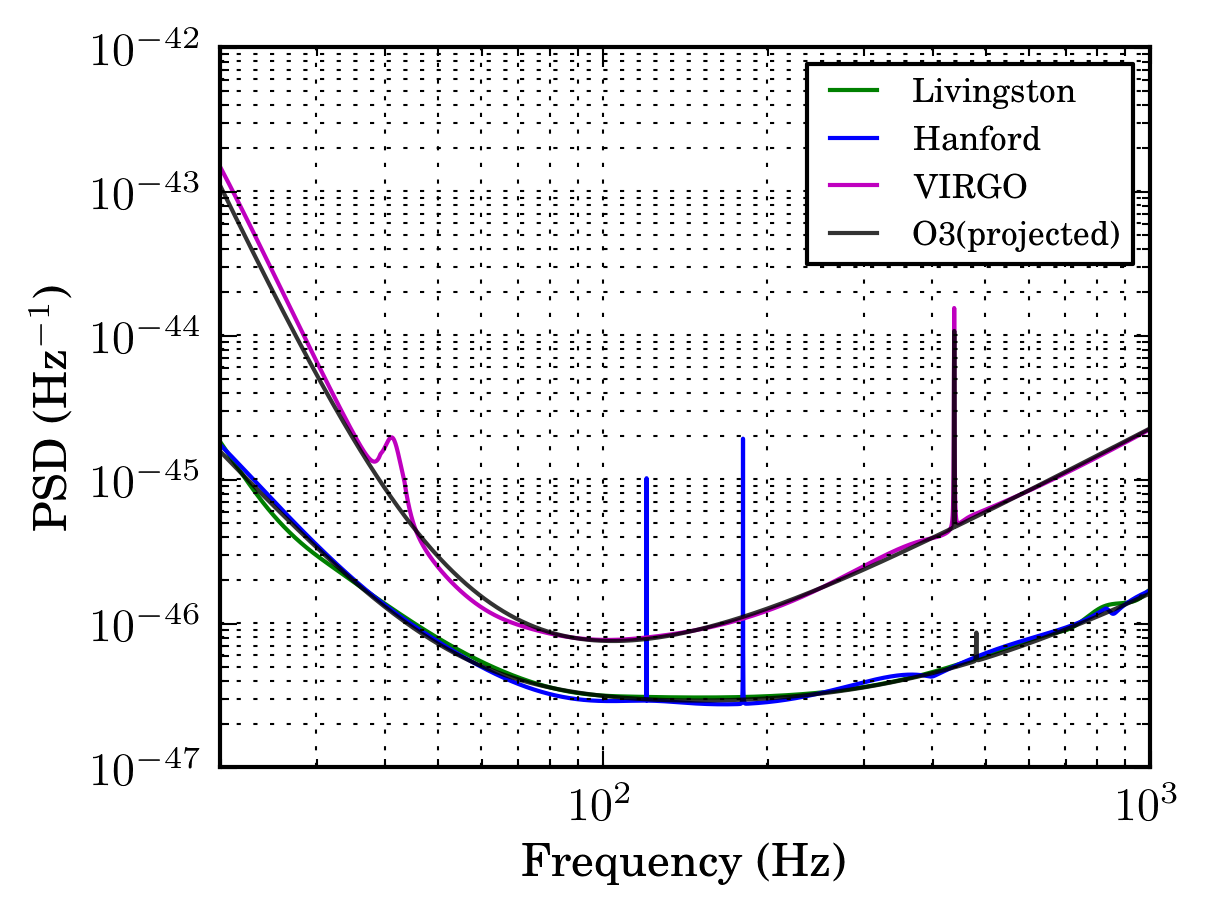}
        \caption{PSD of the recolored data for advanced LIGO (green for Livingston, blue for Hanford) and advanced Virgo (purple). The black lines are PSD for each instrument at the projected O3 sensitivities~\cite{ligosen}.}
        \label{PSD}
\end{figure}

The PSDs \emph{estimated} with the BayesWave algorithm~\cite{2015CQGra..32m5012C} for one of the data streams produced with this method are shown for each interferometer in Fig. ~\ref{PSD} (colored curves), together with the projected O3 curves (black lines) that we use to recolor the O1 data.

\begin{table}[h!]
\centering
\begin{tabularx}{0.45\textwidth}{bZZZb}
 \toprule \toprule
GPS time &  Year & Month & Day & Time  \\ [0.5ex] 
\midrule
 1135924088 & 2016 & Jan & 04 & 06:27:51  \\
 1135989351 & 2016 & Jan & 05 & 00:35:34 \\
 1136267078 & 2016 & Jan & 08 & 05:44:21 \\
 1136506663 & 2016 & Jan & 11 & 00:17:26 \\
 1136594611  & 2016 & Jan & 12 & 00:43:14 \\ [0.5ex] 
\bottomrule\bottomrule
\end{tabularx}
\caption{O1 times (in GPS and GMT) used for the analyses. These are the times at which signals are ``injected''.}
\label{Table.GPS}
\end{table}

We can describe the gravitational waveform emitted by a binary of two point masses in a circular orbit by a set of 15 parameters, $\boldsymbol {\theta}$, including  masses and spins, as well as extrinsic parameters such as luminosity distance, orbital orientation, and sky position~\footnote{BNS would require two additional parameters to model their linear tidal deformability~\cite{2008PhRvD..77b1502F,2014PhRvD..89j3012W,2017PhRvL.119p1101A,Lattimer:2013,2014PhRvD..89b1303Y} In this study we neglect tidal effects for BNS and NSBH to keep the computational cost reasonable. Given that tidal deformabilities are hard to measure even for loud events such as GW170817, they would have been unmeasurable with the marginal sources we consider in this work.}.
Different parametrizations are possible for the mass parameters. In this work we will report the uncertainty of the asymmetric mass ratio $q = m_2/m_1$ ($m_1$, $m_2$ are the component masses, with $m_1 \geqslant m_2$ by convention), and the chirp mass $\mathcal{M}$:
\begin{equation}
\mathcal{M} = (m_1m_2)^{3/5}(m_1+m_2)^{-1/5}
\label{Eq.mc}
\end{equation}

In general, six parameters are needed to describe the spins of the binary: 2 dimensionless spin magnitudes $a_1$, $a_2$ defined as $a_i = |\mathbf{s_i}|/m_i^2$ where $\bf{s_i}$ is the spin vector; the tilt angles $t_1$, $t_2$ between the spin vectors and the orbital angular momentum at some reference frequency (equal to the lower frequency used in the analysis, see below) and two azimuth angles.	

To generate the waveform signals, we use the IMRPhenomPv2 waveform family. This is an inspiral-merger-ringdown (IMR) waveform with an effective precessing spin~\cite{PhysRevLett.106.241101,2013PhRvD..88f2001A,Hannam:2013oca}. We keep all the phase and amplitude corrections supported by the waveform. 

The loudness of a given waveform can be assessed with its optimal signal-to-noise ratio $\rho$, defined in each interferometer by the inner product of a waveform with itself~\cite{1994PhRvD..49.2658C}:
\begin{equation}
\label{Eq.SNR}
\rho^2 =\langle h(\boldsymbol {\theta}, f)|h(\boldsymbol {\theta}, f) \rangle \equiv 4\int^{f_\mathrm{high}}_{f_\mathrm{low}}\frac{h(\boldsymbol {\theta},f) h(\boldsymbol {\theta},f)^*}{S_n(f)}df
\end{equation}

\noindent where $h(\boldsymbol {\theta},f)$ is the frequency-domain waveform projected in the detector, and ${S_n(f)}$ is the PSD of the detector noise. 

We stress that the optimal SNR is not exactly the output of search algorithms, which instead calculate a matched-filter SNR (defined as the inner product of the best waveform template with the data~\cite{SathyaSchutzLRR}).  We prefer to work with the optimal SNR since that can be calculated from the waveform being added to the noise, without any knowledge of the exact realization of the noise. The optimal SNR is the expectation of the matched-filter SNR in the limit where the noise can be considered Gaussian. 
In the rest of the paper, we will just refer to optimal SNR as SNR, unless otherwise specified.

Clear detections typically have matched-filter network SNRs well above 12 for heavy objects~\cite{O1-BBH, GW150914-DETECTION, GW151226-DETECTION, 2017PhRvL.118v1101A, 2017ApJ...851L..35A, 2017PhRvL.119n1101A, LIGOScientific:2018jsj}, whereas BNS can be detected with high confidence at lower matched-filter network SNR~\cite{2017PhRvL.119p1101A,0004-637X-849-2-118}. Since we wish to focus on marginal events, we generate signals with lower network SNRs: $[6,7,8,9,10,12]$. We consider 4 representative CBC systems, BNSs, NSBHs, stellar-mass BBHs, and heavy stellar-mass BBHs (hBBHs).
These representative morphologies capture some of the key features that can be present in CBC signals: long inspirals with little or no spin (BNSs); high mass ratio and visible spin precession (NSBHs) and heavier objects with little inspiral (BBHs and hBBHs). The masses and spins of each system are given in Table~\ref{Table.Injection}. When the inclination angle is larger than $\sim70$ degrees, the cross polarization becomes negligible, affecting the estimation of the extrinsic parameters~\cite{Chen:2018omi,Vitale:2018wlg,Usman:2018imj}. Thus, we consider two different inclination angles for each source: $30$ degrees and $80$ degrees, to take into account the effect of polarization. We assign the same geographical coordinate to all of our simulated sources. Specifically, they are overhead of the two LIGO detectors, which is the position where most events should be detected. This results in roughly equal SNR in the two LIGO detectors. We can quantify the sensitivity of each detector to a particular direction and polarization with the square root of the sum of the antenna pattern, F, squared~\cite{SathyaSchutzLRR}:
$$F^{\mathrm{IFO}}= \sqrt{F^{\mathrm{IFO}}_+(\alpha,\delta,\psi)^2 +F^{\mathrm{IFO}}_\times(\alpha,\delta,\psi)^2 }$$

\noindent where $\alpha$ is the right ascension, $\delta$ is the declination, and $\psi$ is the polarization of the GW signal. 

For all of our sources we have the same values: $F^{H}=F^{L}=0.7$, $F^V=0.3$.
Each system has its distance scaled to give the desired SNR and added to the recolored data at the 5 GPS times of Table~\ref{Table.GPS}. In total, we analyze 4 (mass bins) $\times$ 2 (inclinations) $\times$ 6 (SNRs) $\times$ 5 (GPS times) = 240 simulated events.

 \begin{table}[h!]
\centering
\begin{tabularx}{0.45\textwidth}{bYYYYYY}
\toprule\toprule
Type &  $m_1 (M_\odot)$ & $a_1$  & $t_1 (\degree)$& $m_2 (M_\odot)$  & $a_2$ & $t_2 (\degree)$\\ [0.5ex] 
\midrule
BNS & 1.4 & 0 & 0 & 1.4 & 0 & 0 \\ 
   NSBH & 8 & 0.8 & 46 & 1.4 & 0 & 0\\
  BBH &12 & 0.6 & 60 & 6 & 0.1 & 60  \\
hBBH & 30 & 0.6 & 60 & 30 & 0.6 & 60  \\
\bottomrule\bottomrule
\end{tabularx}
\caption{The intrinsic parameters for the 4 morphologies considered in this study.}
\label{Table.Injection}
\end{table} 

\subsection{Bayesian parameter estimation} 

In this section we  describe the method used to measure the unknown parameters of the detected signals.
Given that Bayesian inference in the field of gravitational waves is now standard (Refs.~\cite{2015PhRvD..91d2003V, GW150914-PARAMESTIM} provide excellent descriptions) we only quickly review it. 

Given the data in the frequency domain $\boldsymbol {d}$ and a signal model hypothesis H, the posterior probability density function (PDF) can be found using Bayes' theorem,

\begin{equation}
\label{Eq.Posteriors}
p({\boldsymbol {\theta}}|{\boldsymbol {d}},H) = \frac{p({\boldsymbol {\theta}}|H)p({\boldsymbol {d}}|{\boldsymbol {\theta}},H)}{p({\boldsymbol {d}}|H)}
\end{equation}
where $p({\boldsymbol {\theta}}|H)$ is the prior probability density of ${\boldsymbol {\theta}}$ given the hypothesis H, and $p({\boldsymbol {d}}|{\boldsymbol {\theta}},H)$ is the likelihood. 
Under the assumptions that the noise streams are statistically uncorrelated, the likelihood can be written as a product of individual likelihood from each interferometer as \cite{BayesianCoherence}
\begin{equation}
p({\boldsymbol {d}}|{\boldsymbol {\theta}},H) =  \prod_{\mathrm{IFO}}p({\boldsymbol {d}^{\mathrm{IFO}}}|{\boldsymbol {\theta}},H)
\end{equation}
where the product spans all the instruments in the network. The normalization constant $p({\boldsymbol {d}}|H)$ is called the evidence of the data, Z:

\begin{equation}
Z = p({\boldsymbol {d}}| H) = \int d\theta_1...d\theta_N p({\boldsymbol {d}}|{\boldsymbol {\theta}},H) p(\boldsymbol{\theta}|H) 
\end{equation}

Given the multidimensional posteriors in Eq.~\ref{Eq.Posteriors}, the posterior PDFs of any specific parameter can be found by marginalizing over all the other parameters:

\begin{equation}
p(\theta_1|{\boldsymbol {d},H) = \int d\theta_2...d\theta_N p(\boldsymbol{\theta}}|{\boldsymbol {d}},H) 
\end{equation}

We use nested sampling implementation of \textit{LALInference}~\cite{2015PhRvD..91d2003V} to stochastically explore the parameter space and produce posterior distributions for $\boldsymbol {\theta}$.
To reduce the computational cost of the likelihood evaluations, we use the reduced order quadrature (ROQ) approximation~\cite{Smith:2016qas}.
Sampling the parameter space to measure the properties of weak signals in presence of strong priors can be challenging. In Appendix~\ref{Sec.AppendixSampling} we report on some sanity checks we have performed to verify the code had properly converged.

For BNS events we start the analysis at  24Hz, following the LIGO and Virgo collaborations~\cite{2017PhRvL.119p1101A}, while for all other sources we start at 20Hz.

\subsection{Choice of priors}\label{Sec.prior}

Bayesian inference requires explicit priors, in our case $p({\boldsymbol {\theta}}|H)$ under the hypothesis that a CBC signal is present in the data.
For all parameters except the luminosity distance (see below), we use the same priors used by the LIGO and Virgo collaborations for the CBCs detected so far~\cite{GW150914-DETECTION, GW150914-PARAMESTIM, GW151226-DETECTION, 2017PhRvL.118v1101A, 2017ApJ...851L..35A, 2017PhRvL.119n1101A,2017PhRvL.119p1101A,Abbott:2018wiz}.
For sky position, orbital orientation and spin orientation, we use isotropic priors. For the dimensionless spin magnitude, we used uniform priors in the range $[0,0.89]$ for black holes~\cite{GW150914-DETECTION, GW150914-PARAMESTIM, GW151226-DETECTION, 2017PhRvL.118v1101A, 2017ApJ...851L..35A, 2017PhRvL.119n1101A} and $[0,0.05]$ for neutron stars~\cite{2017PhRvL.119p1101A,Abbott:2018wiz}. The priors on the component masses are also uniform, in the range in which the corresponding ROQ basis is valid~\cite{Smith:2016qas}.

The most natural prior for the luminosity distance would be a prior uniform in volume, $p(d_L) \sim d_L^2$, since the detection horizon for CBC is tens or hundreds of megaparsecs, depending on the total mass. This is indeed the prior used in LIGO-Virgo papers.

In this study we prefer to use a \emph{uniform} prior on the luminosity distance, and then reweight the samples to enforce a uniform-in-volume prior. The reason is as follows. The nested sampling algorithm samples the prior~\cite{2004AIPC..735..395S} to calculate the evidence Z, and obtains the posterior distribution as a by-product. 
When using a uniform-in-volume distance prior, the nested sampling algorithm will spend a significant fraction of time exploring the region of parameter space where the distance is large.
If we could run the algorithm for an arbitrarily large number of steps, it would eventually converge to the correct parameters. But for low-significance events like the ones we are considering, and using a finite number of steps, the algorithm might in practice not explore properly the part of the parameter space where the distance is small, since it would have to overcome a significant prior penalty.
By sampling with a uniform-in-distance prior we avoid this issue. The sampler can explore with ease the whole distance range, and the correct prior is applied in post-processing with a standard rejection sampling approach.

\section{Results}
\label{Results}

In this section we report the uncertainty in measuring some of the key parameters of the simulated events.
Unless otherwise specified, we used 90\% credible intervals (CI). Those are either absolute intervals (in the appropriate units), or relative to the true value (in percentages).
\subsection{Extrinsic parameters}

Extrinsic parameters such as sky location and luminosity distance are of fundamental importance for EM follow-ups. 
Although the details of the follow-up to GW triggers vary with the facility and geographical factors (e.g. altitude of the source)~\cite{2017ApJ...848L..12A}, one usually tries to cover all or part of the sky uncertainty area (or sky-distance uncertainty volume). In fact, 3-dimensional uncertainty volumes have been routinely released in low latency by the LIGO and Virgo collaborations for significant events in the first and second observing run. Lowering the threshold for making triggers public would in general imply a significant increase of the false positive fraction~\cite{Lynch:2018yom}. In this section, we show that additionally, even for the events which are of astrophysical origin, an EM follow-up campaign might be prohibitive. 

\begin{figure}[htb]
  \centering
    \includegraphics[width=0.45\textwidth]{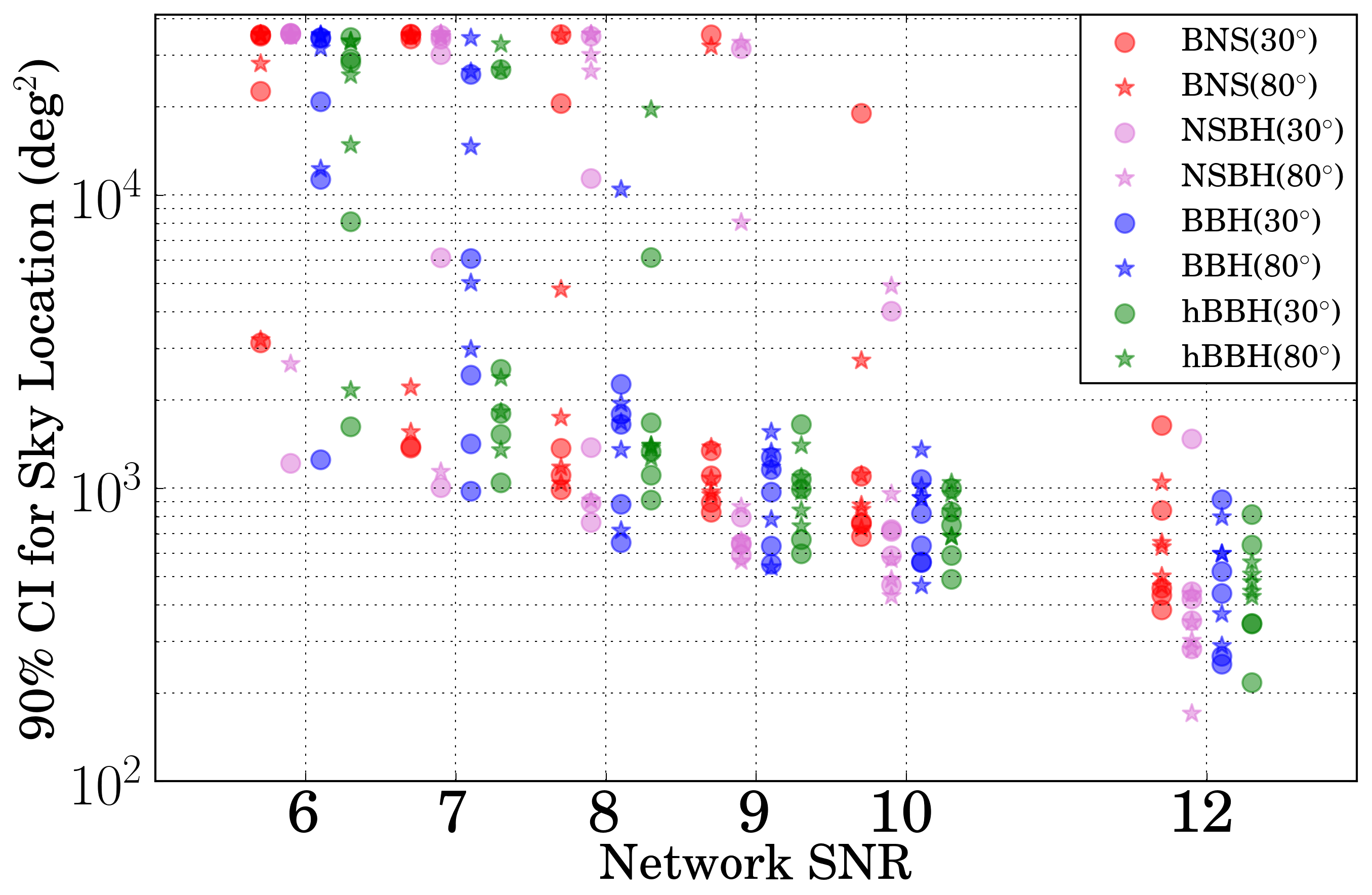}
    \caption{90\% credible interval of the marginalized posteriors of the sky location vs. network SNR. }
  \label{sky_snr}
\end{figure}

In Fig.~\ref{sky_snr}, we show the 90\% credible regions (\degg) for all the sources we analyzed. 
Each color represents a different morphology (red for BNS, purple for NSBH, blue for BBH, green for hBBH), whereas the shape of the bullet indicates the inclination angle (circles for $30^\circ$, stars for $80^\circ$). 
The optimal network SNR (for the remainder of this section we will drop the ``network'') is reported in the horizontal axis. Notice that we have artificially introduced a small horizontal offset while plotting to avoid significant superposition between data from different morphologies. The 5 symbols for each (SNR, morphology, inclination) set are the results from the 5 GPS times.

We see the expected overall trend of decreasing uncertainties with increasing SNRs. At SNRs below 10, we find that some of the sources are localized to uncertainties of $10,000$ \degg or more. 
We have verified that the outliers are due to the specific data stream into which events were added. In particular, we find that 1 of the 5 GPS times produces systematically lower uncertainties. At these low SNRs, the actual noise realization can have significant impact on the outcomes of the parameter estimation process.
As the SNRs increase, the signals can be more easily distinguished from the noise, and the latter plays less of a role.

The loudest events we consider have SNRs of 12. For most of those, the 90\% credible regions of sky localization are of [200-1,000] \degg. Even at SNRs below 10, there are sources that can be localized within areas of {$[400-1,000]$ \degg}. 
For comparison, Ref.~\cite{2014arXiv1404.5623S} simulated a large number of BNS with astrophysically motivated parameters, and found that 50\% of the BNS detectable by a LIGO-Virgo network (at their O2 sensitivity) would have 90\% CI of {235}~\degg or smaller if a detection threshold of matched-filter network SNR above 12 is used.
While follow-up will be challenging, these large error areas may be accessible by wide-field/all-sky survey instruments available across the electromagnetic band and neutrinos. Here below we  discuss some of the instruments. 

The most viable option would be radio~\cite{2016ApJ...831..190H,NakarNature,2016ApJ...829L..28P}, where wide-field instruments like the Karl G. Jansky Very Large Array and the Very Long Baseline Array~\cite{1994IAUS..158..117N} of the National Radio Astronomy Observatory, and the Long Wavelength Array~\cite{5109716} can cover a large fraction of the sky. Coincident, high-energy (GeV to EeV) neutrinos might be searched in all-sky neutrino observatories like the IceCube~\cite{Aartsen:2016nxy} or the Antares~\cite{2011NIMPA.656...11A}. 
Gamma-ray bursts can be found independently of the GW observation, which could be used to confirm the astrophysical nature of the GW candidate. The Fermi Gamma-ray Burst Monitor~\cite{0004-637X-702-1-791} aims to localize strong burst near the center of the field of view (FoV) of the Large Area Telescope, which is over 6,000 \degg. From Earth, the High-Altitude Water Cherenkov~\cite{0004-637X-843-1-39} has an instantaneous FoV of over 6,000 \degg (15\% of the sky). 
However, given the large uncertainties in sky position and distance, one might have to deal with a significant background of EM signals, which would make it difficult to claim a solid association.  
In this scenario, requiring both position and time coincidence (for the prompt EM emission) might increase the chances of success. Even at SNRs of 6, the arrival time of the GW signals at the detectors can be determined within a fraction of a second, see Appendix~\ref{Sec.AppendixTime}. 
Instruments working at optical frequencies usually have smaller FoV: for example, the Asteroid Terrestrial-impact Last Alert System~\cite{1538-3873-130-988-064505} has a FoV of 60 \degg, but it scans a large fraction of the sky every night of operation.
There might be more observatories available at the time of O3. For example, the Zwicky Transient Facility~\cite{2014SPIE.9147E..79S,2016SPIE.9908E..5MD} with a FoV of 47 \degg in the optical, and the Cherenkov Telescope Array~\cite{2011ExA....32..193A} with whole sky coverage for gamma rays. Beyond O3, the Large Synoptic Survey Telescope~\cite{2016arXiv161001661N} with a FoV of 9.62 \degg in the optical might come online in 2023.

\begin{figure}[htb]
  \centering
    \includegraphics[width=0.45\textwidth]{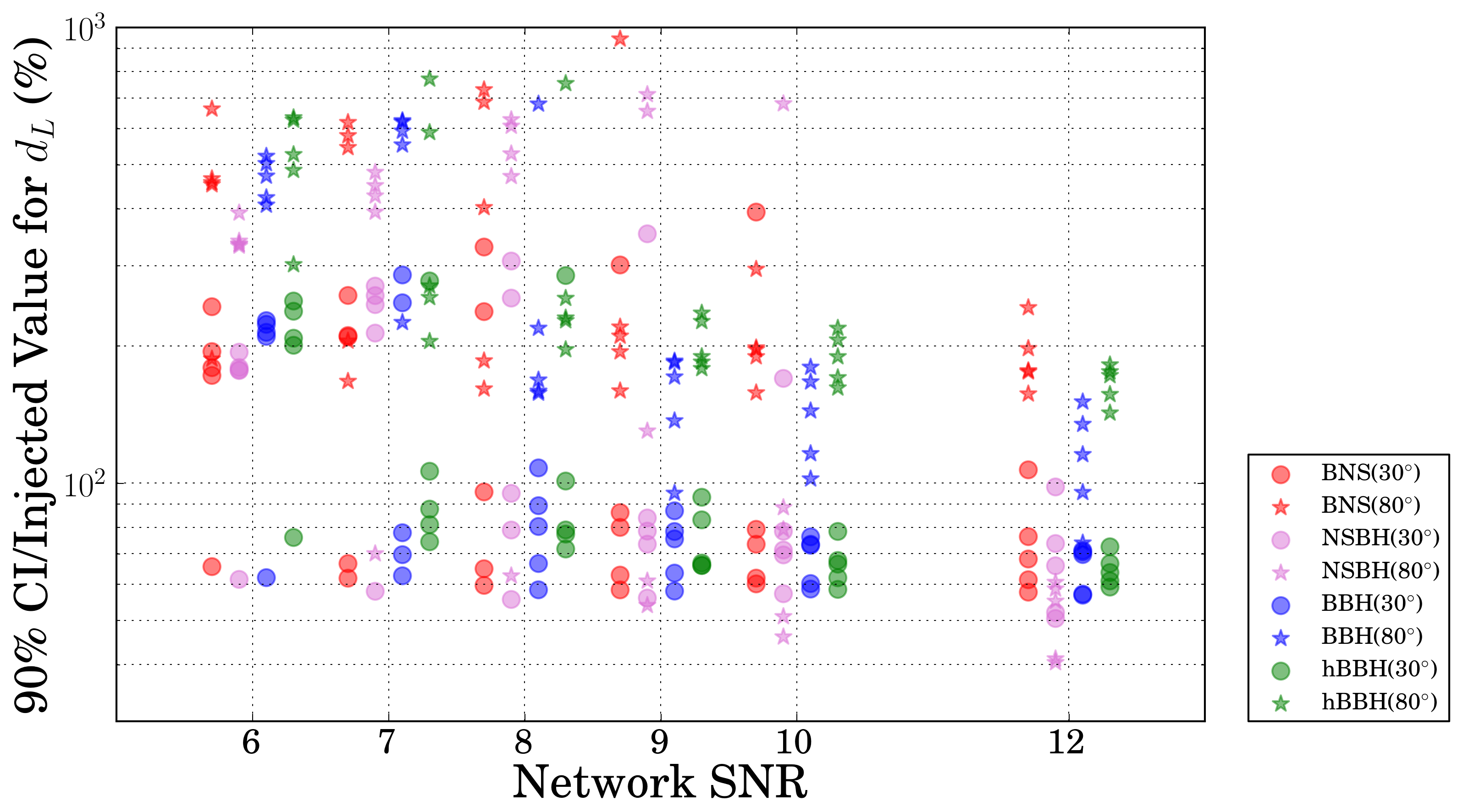}
    \caption{Relative 90\% credible interval of the marginalized posteriors of the luminosity distance $d_L$ vs. network SNR. }
  \label{dist_snr}
\end{figure}

Similar large variations in precision can be seen for the luminosity distance $d_L$, Fig.~\ref{dist_snr}.
Relative uncertainties are above 40\% at all SNRs, and very often above 100\%. 
We observe that NSBH systems typically yield the smallest uncertainties. This is expected and is due to the fact that we allowed spin precession for NSBH systems, as given in Table.~\ref{Table.Injection}. Spin precession, together with the high mass ratios of NSBH, helps breaking the degeneracy between distance and inclination angle, leading to smaller uncertainties~\cite{PhysRevD.49.6274,Vitale:2018wlg, PhysRevLett.112.251101}.  

\subsection{Intrinsic parameters}

We now discuss the estimation of intrinsic parameters: mass and spins.

For clear detections, the detector-frame chirp mass [defined in Eq.~\eqref{Eq.mc}] is typically measured very well, with relative uncertainties of $\sim0.01\%$ for BNS sources~\cite{Abbott:2018wiz, 2016ApJ...825..116F,VeitchMandel:2012} and around 15\% for heavier objects~\cite{O1-BBH, GW150914-DETECTION, GW150914-PARAMESTIM, GW151226-DETECTION, 2017PhRvL.118v1101A, 2017ApJ...851L..35A, 2017PhRvL.119n1101A,LIGOScientific:2018jsj}. 
The reason why the chirp mass is measured better for low-mass systems is trivially that it affects the phase evolution at the lowest post-Newtonian order~\cite{2006LRR.....9....4B}, and thus can be measured better for signals with long inspirals.
Since the merger frequency decreases as the total mass increases, BNSs are the sources for which the chirp mass can be best measured.

\begin{figure}[htb]
  \centering
    \includegraphics[width=0.45\textwidth]{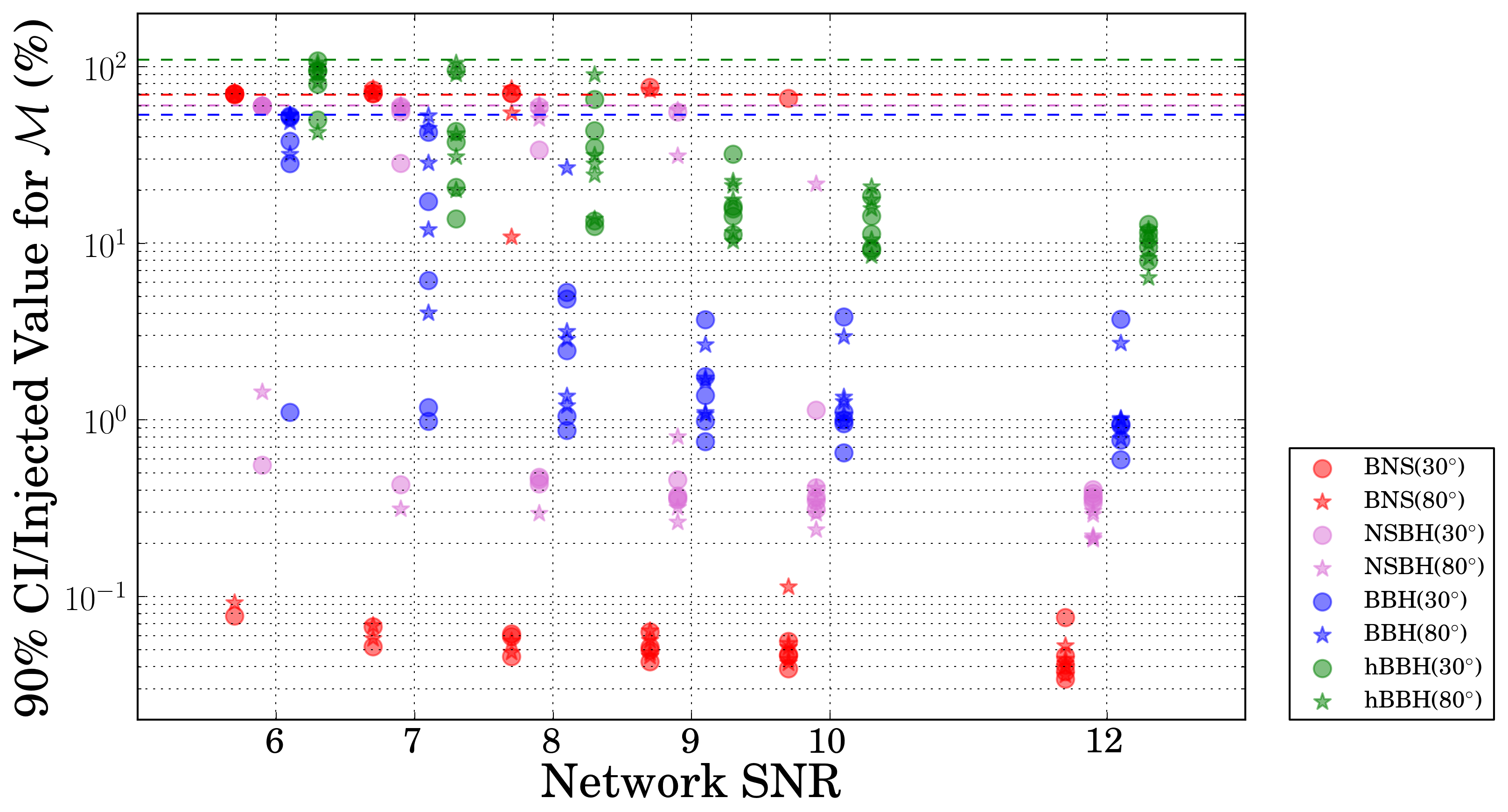}
    \caption{Relative 90\% credible interval of the marginalized posteriors of the detector-frame chirp mass $\mathcal{M}$ vs. network SNR. The dashed lines represent the relative 90$\%$ width of the prior.}
  \label{Fig.mc_snr}
\end{figure}

In Fig.~\ref{Fig.mc_snr}, we report the relative 90\% uncertainty for the chirp mass against the network SNR.
Overall, we see that BNS have the smallest relative uncertainties followed by NSBH, BBH and heavy BBH. The different morphologies are thus naturally sorted by total mass, as expected. 

Let us first discuss the high-SNR end of our simulations, as the weakest events deserve a separate discussion. 
At SNR 10 and 12, the BNSs have relative uncertainties between $0.03\%$ and $0.1\%$.
This is in the ballpark of what one could have guessed using a simple 1/SNR$^2$ scaling argument~\cite{1994PhRvD..49.2658C}, and GW170817 as a reference point.
That simple analysis is not totally accurate, since the variance scales like 1/SNR$^2$ only at large SNRs. For smaller SNRs, higher order corrections must be taken into account~\cite{2010PhRvD..82l4065V,2010PhRvD..81l4048Z}, and the uncertainties in the plot are indeed well above what a simple scaling argument would suggest. 
At the high end of our SNR distribution we find that, with very few exceptions, the results from different GPS times and orientations are grouped together, and the 
uncertainty mostly depends on the morphology. 
Due to their relatively long inspirals, NSBH are the second-best type of source, with uncertainties between $0.2\%$ and $\sim2\%$ at SNRs 10 and 12, including all data realizations and orientations. Stellar-mass black hole binaries have uncertainties between $0.5\%$ and $\sim4\%$, a factor of 2 worse than NSBH, while heavy BBH are considerably worse, with uncertainties between $5\%$ and $30\%$ (again, at SNR 10 and 12).

As the SNRs decrease, we observe 2 distinct populations: one that roughly continues the trend we see at SNR 10 and 12 with some degradation; and another with much larger uncertainties, comparable to the prior width, especially for BNSs.
The latter are events for which the posteriors are not unimodal. One could expect that at low SNRs, noise fluctuations can seriously impact the measurement of the chirp mass, since this is obtained by ``following'' the phase evolution of the waveform signal through thousands of cycles~\cite{2005PhRvD..71h4008A}.
We get a (rough) classification of the chirp mass posteriors by using the \textit{find\_peaks} routine of Scipy~\cite{findPeaks} using a prominence threshold of 9\% of the main peak to trigger the presence of additional modes. While more sophisticated statistical tests could be used,  this gives us at least an idea of which fraction of posteriors is clearly unimodal. A distinct type of chirp mass posterior appears often enough to deserve some discussion: those with clearly unimodal posteriors but with tails spanning large enough to cover most or all of the prior range. These are sources for which the likelihood profile is shallow, and does not dominate the underlying prior distribution.

\begin{figure}[htb]
    \centering
    \begin{subfigure}[b]{0.3\textwidth}
        \includegraphics[width=\textwidth]{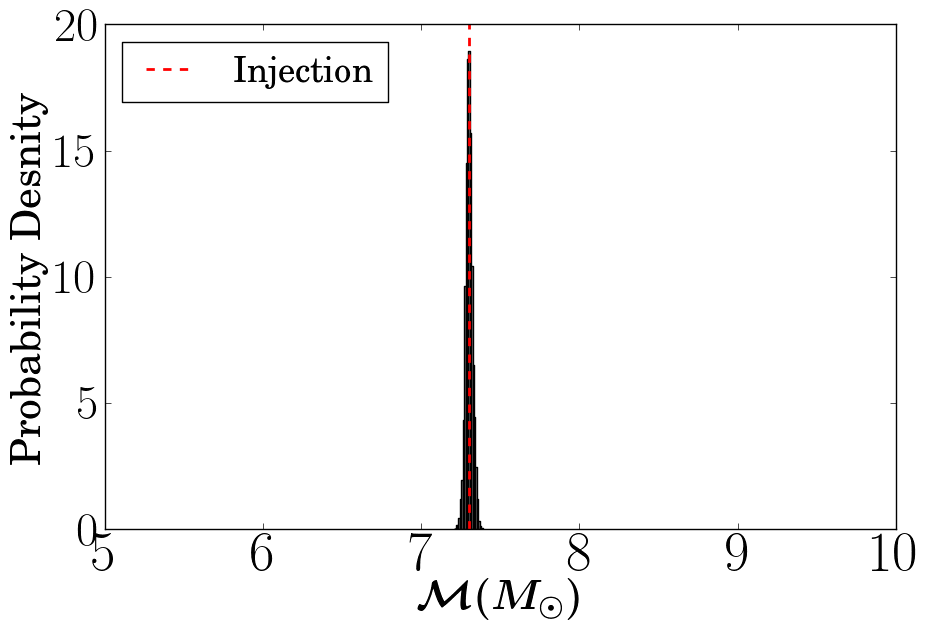}
        \caption{Unimodal. The network SNR is 12. }
        \label{uni}
    \end{subfigure}
    ~ %
    \begin{subfigure}[b]{0.3\textwidth}
        \includegraphics[width=\textwidth]{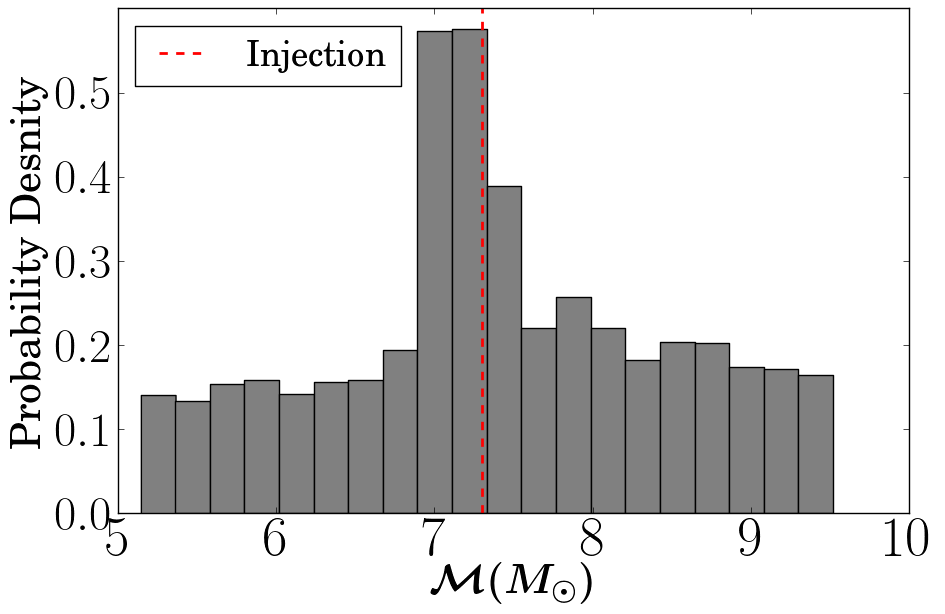}
        \caption{``Unimodal-Wide''. The network SNR is 7. }
        \label{uni-wide}
    \end{subfigure}
    ~ %
    \begin{subfigure}[b]{0.3\textwidth}
        \includegraphics[width=\textwidth]{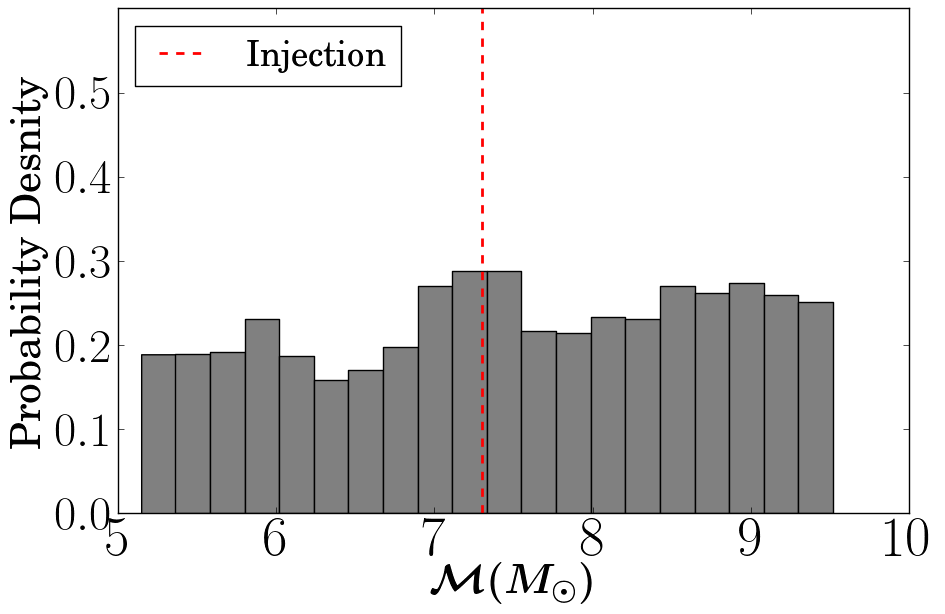}
        \caption{Multimodal. The network SNR is 6. }
        \label{multi}
    \end{subfigure}
    \caption{Representative posterior distribution for $\mathcal{M}$. All systems are BBHs with true inclination angle of 80$\degree$. The range is the same in all panels in the x axis.}
    \label{Fig.PosteriorExamples}
\end{figure}

We thus introduce 3 different categories: unimodal; ``unimodal-wide'', which we define as posteriors with one clear peak but a 90\% uncertainty larger than 50\% of the prior range; and multimodal posteriors. Examples of each are shown in  Fig.~\ref{Fig.PosteriorExamples}.

\begin{figure}[htb]
  \centering
    \includegraphics[width=0.4\textwidth]{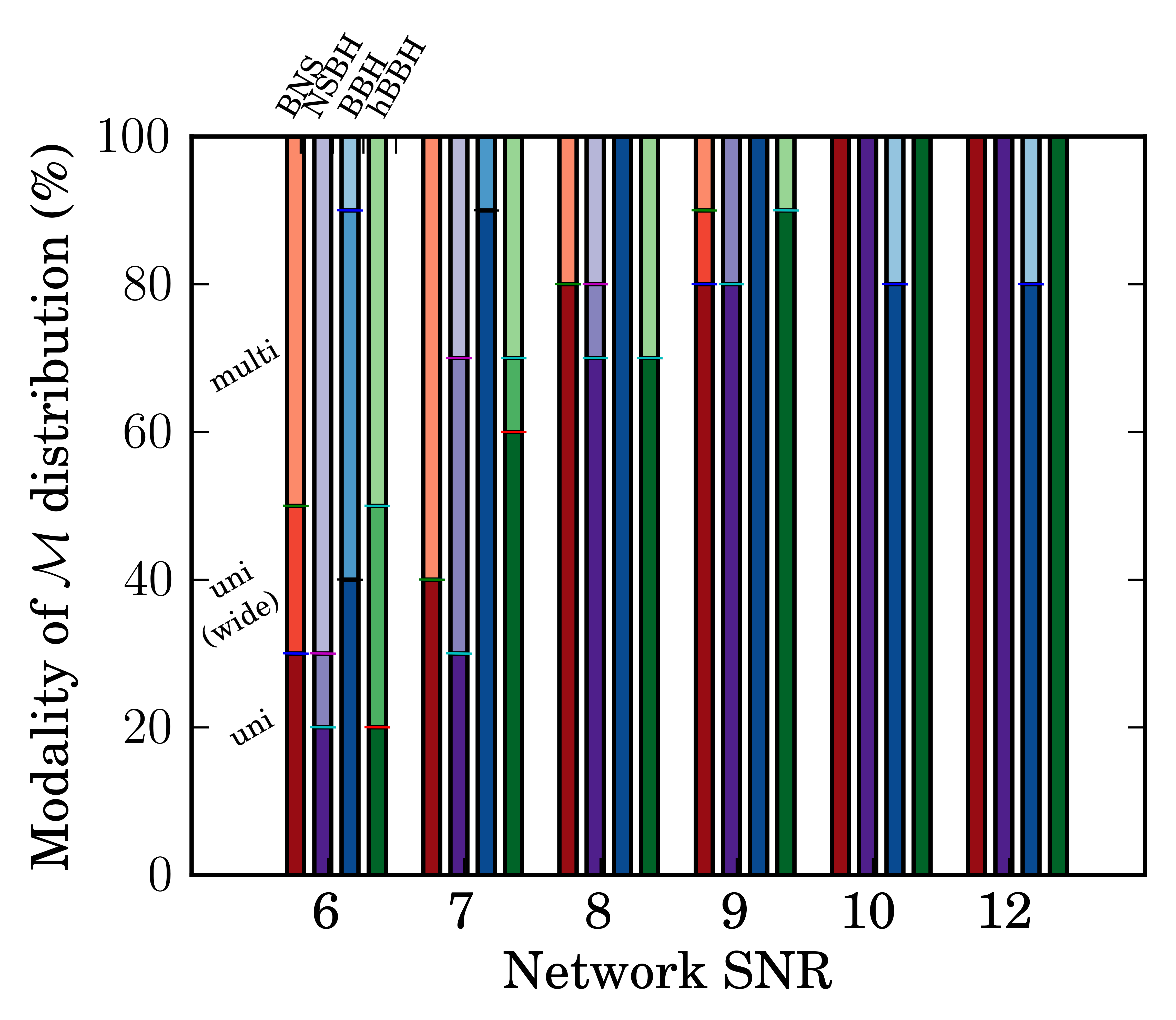}
    \caption{Proportions for unimodal, unimodal (wide) and multimodal posterior distributions for detector-frame chirp mass $\mathcal{M}$ (\%) vs. network SNR. }
  \label{Fig.nmodes}
\end{figure}

We apply this classification scheme to all events, and show the fraction of events that belong to each category in Fig.~\ref{Fig.nmodes}. At SNR 10 and 12, all morphologies but BBH show unimodal posteriors. Some of the BBH sources at these SNRs have hints of secondary modes which, however, do not significantly broaden the 90\% credible intervals, as clear from Fig.~\ref{Fig.mc_snr}. 
On the opposite end, at SNRs of 6 and 7, most events have multimodal posteriors, or a broad posterior filling up most of the prior range with just a hint of peak at the true chirp mass value.

While we directly measure detector-frame chirp mass, the more astrophysically interesting parameter is source-frame chirp mass \mcsource: 
$$\mathcal{M} = (1 + z) \mathcal{M}^{\mathrm{source}}$$

\begin{figure}[htb]
  \centering
    \includegraphics[width=0.45\textwidth]{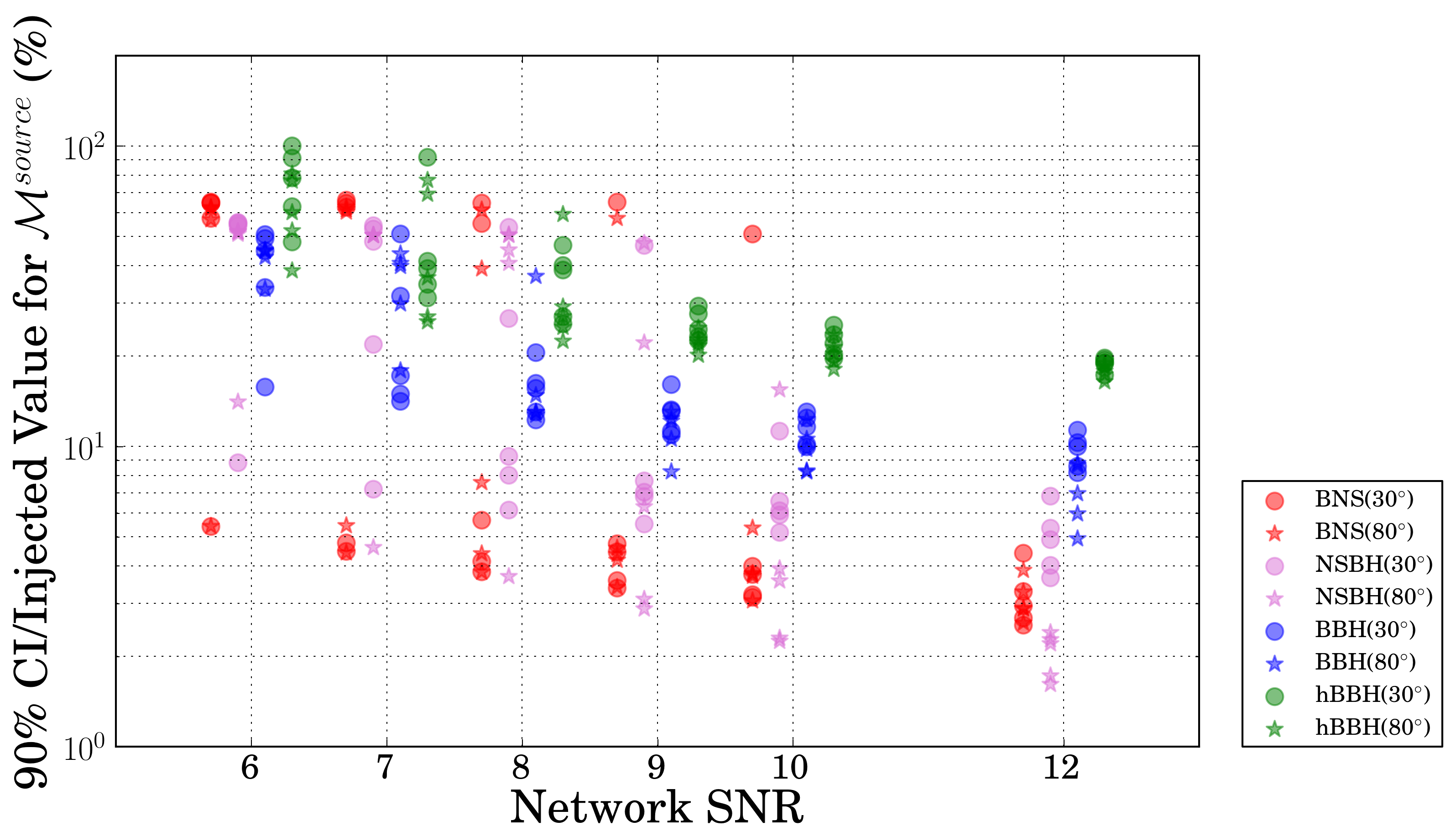}
    \caption{Relative 90\% credible interval of the marginalized posteriors of the source-frame chirp mass \mcsource vs. network SNR. }
  \label{Fig.mc_source}
\end{figure}

Since GWs do not provide a direct  measurement of the redshift, one must use the measured luminosity distance and a fiducial cosmology to convert that into a redshift~\cite{2017PhRvL.119p1101A,Abbott:2018wiz}. Since the uncertainty in the distance are usually large for the events we simulated, the source-frame chirp mass will be measured more poorly than the detector-frame chirp mass, Fig.~\ref{Fig.mc_source}. We note that even though the uncertainties follow similar patterns, the scale is different than what was found in Fig.~\ref{Fig.mc_snr}. The best measured systems of BNSs at SNR 6, for example, have relative uncertainties $\sim 5~\%$ instead of  $0.1~\%$ or better. At SNRs 10 and 12, BNSs are measured to uncertainties $\sim [2 - 5]~\%$, NSBHs $\sim [2 - 7]~\%$, BBHs $\sim [5 - 10]~\%$, and hBBHs above 10 $\%$. Yet, it would be possible to get a better redshift measurement, thus a better source-frame chirp mass measurement, if the host galaxy is identified by EM follow-up campaigns for BNSs or NSBHs.

\begin{figure}[h]
  \centering
    \includegraphics[width=0.45\textwidth]{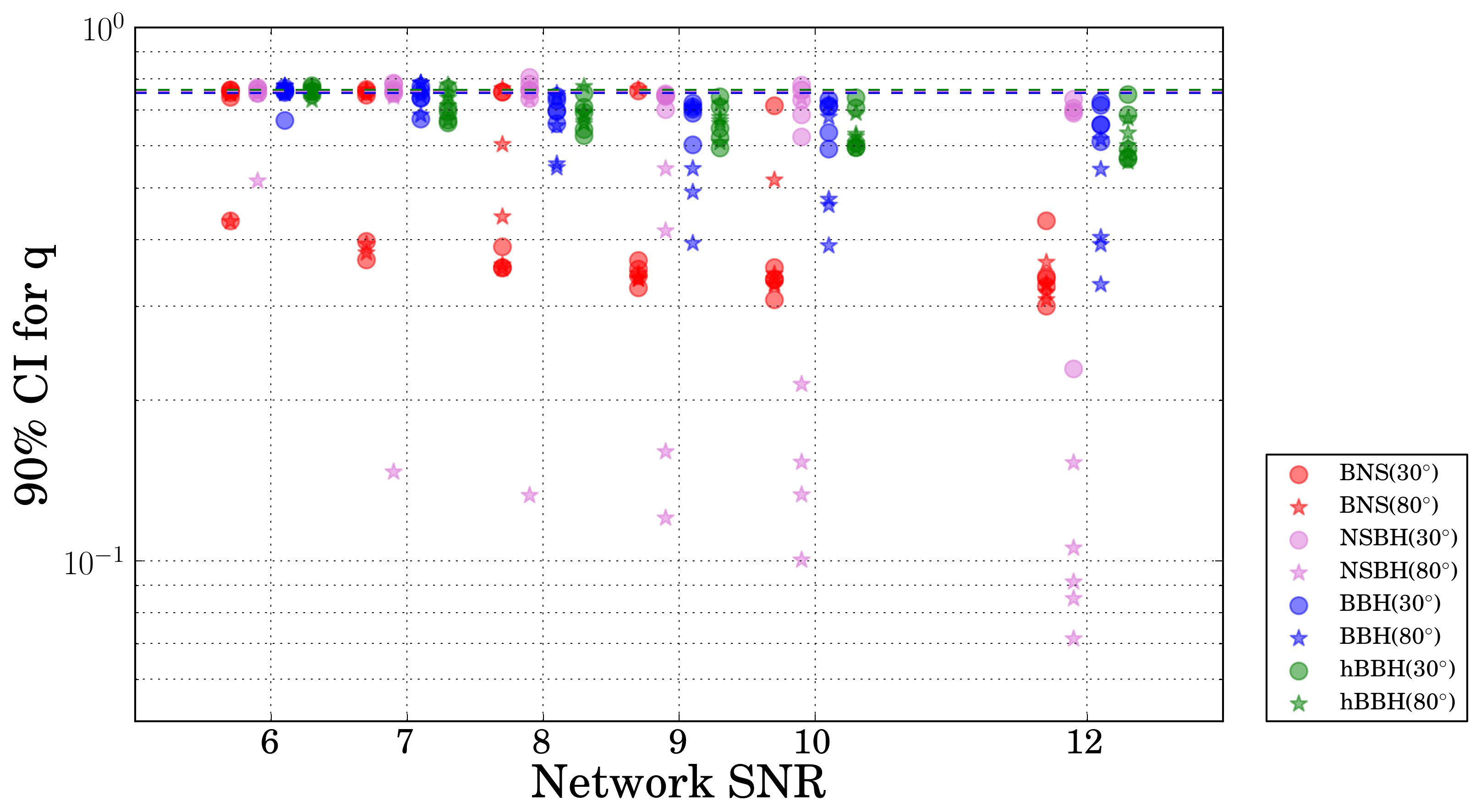}
    \caption{90\% credible interval of the marginalized posteriors of the mass ratio q vs. network SNR. The dashed line represents the 90$\%$ CI for q prior. }
  \label{Fig.mq_snr}
\end{figure}

Next, we consider the asymmetric mass ratio $q$, Fig.~\ref{Fig.mq_snr}.
We expect NSBH to yield the best measurement, since spin-induced precession breaks the degeneracy between mass ratio and spins~\cite{PhysRevD.93.084042,PhysRevD.49.6274,Vitale:2018wlg, PhysRevLett.112.251101}, improving the measurability of both. For NSBH, we find uncertainties as small as 0.07 for SNR 12 and orientations close to edge-on. For the NSBH with orientation close to face-on, the uncertainty is systematically worse, at all SNRs. 
The reason is again correlations. When orbital precession is present, inclinations closer to $90^\circ$ yield better spin measurement~\cite{Vitale:2016avz,PhysRevLett.112.251101}. The smaller spin uncertainties result in smaller mass ratio uncertainties. 
The same trend is visible for the BBHs. However, hBBHs have comparable uncertainties regardless of the orientation, because the hBBHs in our simulations are equal-mass, which suppresses spin precession. 
Owing to their longer inspiral phase, the uncertainties for the BNS are of $[0.3 - 0.5]$ for most BNSs in the SNR range $7-12$, better than for BBHs and hBBHs.
These results suggest that, if the real source is a marginal a NSBH with visible spin precession, one might be able to distinguish between a NSBH and a low-mass BBH , but not so easily between a BNS and a low-mass NSBH. 

We now look at the estimation of the spins. Since individual spins are hard to measure even for loud sources~\cite{O1-BBH,GW150914-DETECTION, GW150914-PARAMESTIM, GW151226-DETECTION, 2017PhRvL.118v1101A, 2017ApJ...851L..35A, 2017PhRvL.119n1101A,Abbott:2018wiz,PhysRevD.93.084042,Vitale:2016avz,PhysRevLett.112.251101,LIGOScientific:2018jsj} we focus on the effective inspiral spin, \chieff:
\begin{equation}
\chieff= \frac{c}{G(m_1+m_2)}\left(\frac{\boldsymbol{S}_1}{m_1}+\frac{\boldsymbol{S}_2}{m_2}\right)\cdot\frac{\boldsymbol{L}}{|\boldsymbol{L}|}
\end{equation}
where $\boldsymbol{L}$ is the orbital angular momentum. \chieff is the mass-weighted projection of the total spin onto the orbital angular momentum and takes values between -1 and 1~\cite{PhysRevD.64.124013,PhysRevD.78.044021,PhysRevD.82.064016,2011PhRvL.106x1101A}.  While individual spins will not often be measurable, the effective spin is usually measured well~\cite{O1-BBH,GW150914-DETECTION, GW150914-PARAMESTIM, GW151226-DETECTION, 2017PhRvL.118v1101A, 2017ApJ...851L..35A, 2017PhRvL.119n1101A, LIGOScientific:2018jsj, Abbott:2018wiz,Ng:2018neg,Vitale:2016avz}. Importantly, the effective spin can also be used to distinguish between different astrophysical formation channels~\cite{2016ApJ...825..116F,2017PhRvD..96b3012T}, instead of relying on the measurement of the individual spins~\cite{Vitale:2015tea,PhysRevD.87.104028}.

\begin{figure}[htb]
  \centering
    \includegraphics[width=0.45\textwidth]{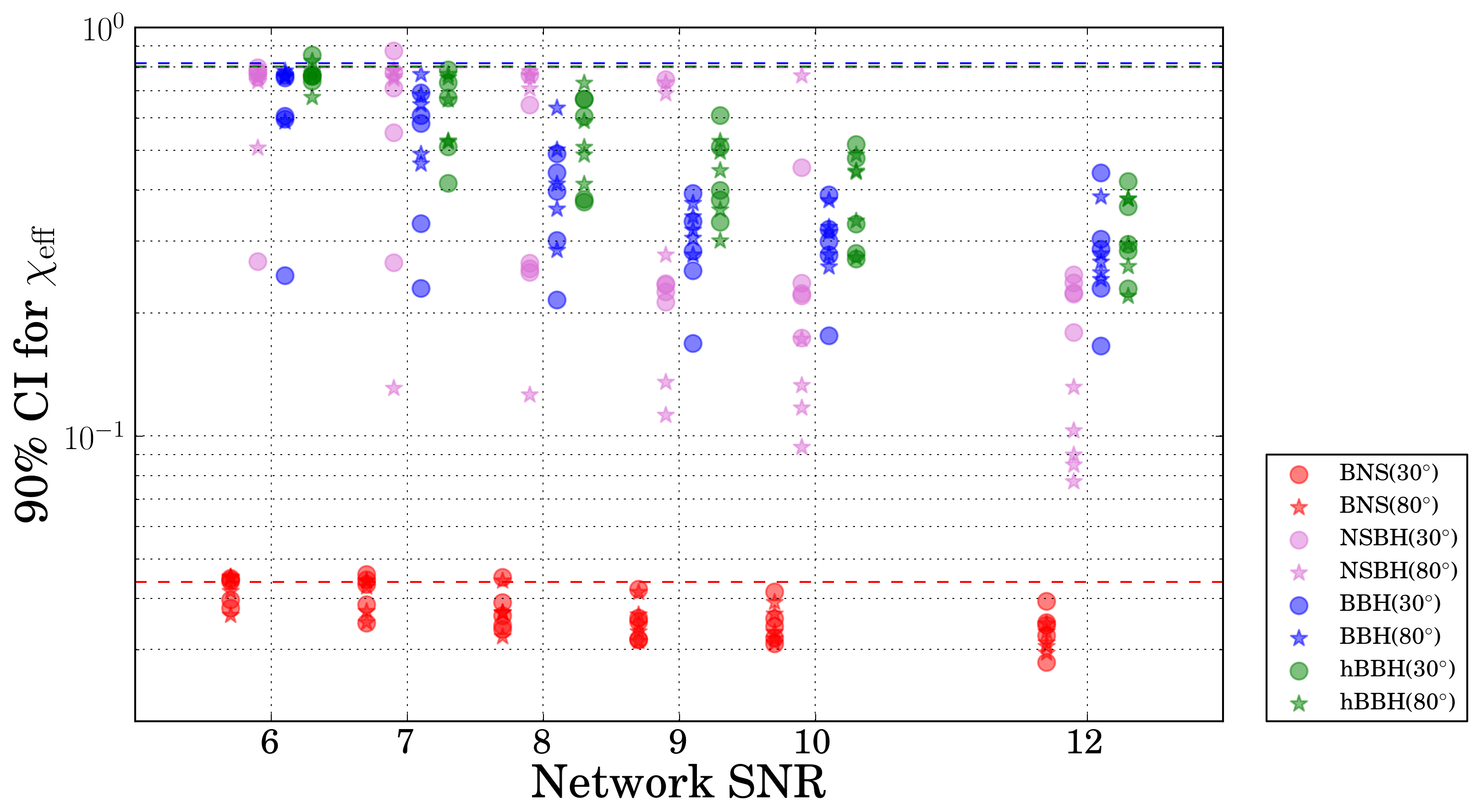}
    \caption{ 90\% credible interval of the marginalized posteriors of the effective inspiral spin \chieff vs. network SNR. The dashed line represents the 90$\%$ CI for \chieff prior. }
  \label{chieff_snr}
\end{figure}

We report the 90\% CI for \chieff in Fig.~\ref{chieff_snr}, where the horizontal dashed lines represent the 90\% of the prior width.
For all sources that include at least one black hole, the uncertainties  at SNR 12 are between $\sim 0.08$ and $\sim 0.3$, that is, a factor of $2-10$ narrower than the prior. At SNR 10, the uncertainties for these systems are between $\sim 0.1$ and $\sim 0.4$.  For the same reasons that we described while discussing the mass ratio results, NSBH oriented edge-on are the events for  which the effective spin estimation is best. Heavy BBH are the worst since they have short inspirals, and mass ratio of unity. As the SNRs decrease, for some sources the \chieff posterior is still informative. Conversely, for the BNS sources, even at SNR 12 the posterior is only marginally narrower than the prior. 

\begin{figure}[htb]
  \centering
    \includegraphics[width=0.45\textwidth]{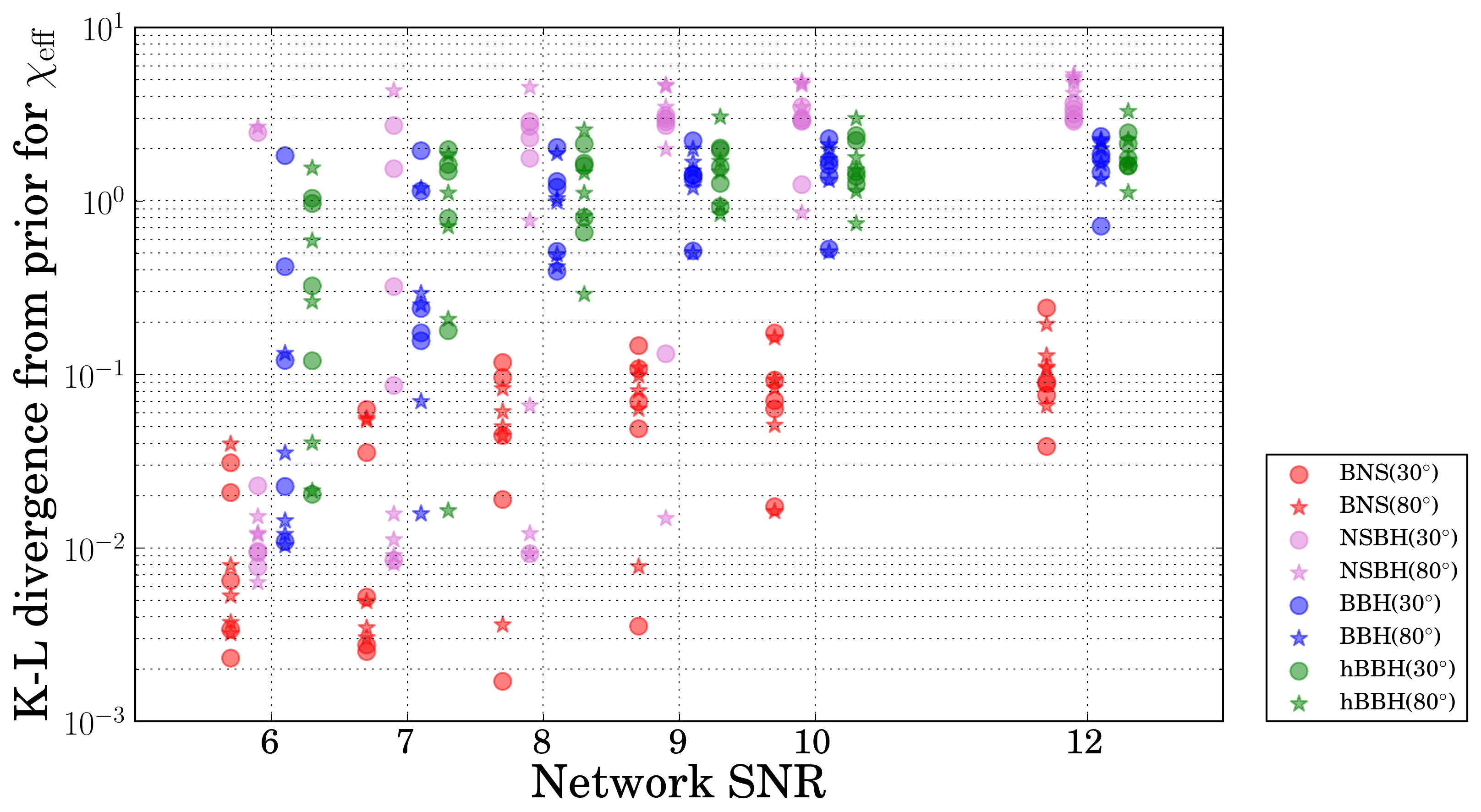}
    \caption{K-L divergence for \chieff vs. network SNR.}
  \label{chieff_KLD}
\end{figure}

To quantify the amount of information gained about \chieff after analyzing the data, we calculate the Kullback-Leibler (K-L) divergence~\cite{klDivergence} of the posterior Q over the prior P:

\begin{equation}
D_{KL}(P||Q) = \sum_i P(i)\mathrm{ln}\frac{Q(i)}{P(i)}
\end{equation}

where the index $i$ spans the samples. This is shown in Fig.~\ref{chieff_KLD}. For most of the sources with at least one black hole at SNR 12, the data yield more than 1 nat of information. NSBHs are the systems for which most information is gained. 
The data is usually less informative for BNSs. At SNR 12, the typical information gained for BNSs is $\sim 0.1$ nats, while at low SNR no information is gained, which implies the posterior is basically the prior. This is consistent with Fig.~\ref{chieff_snr}. As the SNR decreases, the data is not allowed to significantly update the  prior, and the K-L divergence can get values below 0.1 nats.

The effective spin parameter related to precession, \chip~\cite{Hannam:2013oca,2015PhRvD..91b4043S}, is not measurable at these SNRs. Only for NSBH and at high inclinations the data provides some information about \chip. In Appendix~\ref{Sec.K-L Divergence}, we report the median K-L divergence over the 5 GPS times for all parameters and all signal morphologies.

\section{Conclusions}\label{Sec.Conclusions}

Compact binary coalescences, the most common source of gravitational waves detectable by ground-based detectors, are expected to be distributed uniformly in volume. This implies that their SNR $\rho$ should be distributed as $\rho^{-4}$: for each loud detection there should be many more marginal signals in the data. 

In practice, lowering the threshold matched-filter network SNR (or other detection statistics) will not increase the number of detections with the fourth power, since the background from instrumental and environmental sources increases more steeply. As the detection threshold is lowered, candidate events will be competing with background events of non-astrophysical origin~\cite{0004-637X-849-2-118,CBCVeto,1809.03815,Lynch:2018yom}. 

It is still the case, though, that a significant fraction of detections made in the next few years will be marginal. In this paper we have explored what kind of astrophysical information can be extracted from these weak signals. This is a topic that was not extensively explored in the literature, since most existing work focuses on clear detections (but see Ref.~\cite{1809.03815}).

We have simulated GWs from various CBC sources with different network SNRs, from 6 to 12, and added them into real interferometric data. 
We have considered a network made by the two advanced LIGOs and the advanced Virgo detector at their expected sensitivity for the third observing run (planned for early 2019~\cite{ligosen}).

We have shown that the 90\% credible regions in the sky localization of the sources are of $[200-1,000]$~\degg for network SNR of 12, for all signal morphologies. As the network SNR decreases, the uncertainty increases, and a larger spread between different noise realizations is present, which confirms that the specific noise realization can seriously affect the outcome of the analysis at very low SNRs.
At the lowest SNRs we consider, most sources are simply not localized. It is important to notice that for those events it is not systematically the case that other parameters are not measured (e.g., the chirp mass).
The reason is that a single detector is enough to measure an intrinsic parameter like the chirp mass, while at least two are needed to constrain the sky position, through time triangulation as well as amplitude and phase consistency~\cite{Fairhurst2009, 2014ApJ...795..105S,2014PhRvD..89h4060S}.

The measurement of the luminosity distance, which could be used to infer the Hubble constant if a counterpart is found~\cite{2005ApJ...629...15H,2013arXiv1307.2638N,HubbleConstant,Vitale:2018wlg,2017arXiv171206531C} (or statistically, in absence of counterparts~\cite{SchutzNature,2012PhRvD..86d3011D}), is also challenging for similar reasons.
We have found relative uncertainties for the luminosity distance above $40\%$ even for the loudest event we considered, and above $100\%$ for a significant fraction of events. The large sky areas within which these marginal events are localized could be searched by wide-field/all-sky survey instruments available across the electromagnetic band and neutrinos. 
Requiring time coincidence between the gravitational and the prompt EM emission might further increase the significance of a joint detection.

We have verified that intrinsic parameters, such as detector-frame chirp mass and the effective spin, can usually be constrained, and do not simply yield the prior. We have found that the detector-frame chirp mass can be estimated to be better than 0.1\% for BNS, 2\% for NSBH and 4\% for stellar-mass BBHs, at SNR of 10 and 12. As the SNR decreases, the uncertainties increases gradually for the bulk of source, but for some noise realizations the posteriors start being multimodal, or develops fat tails, which dramatically increases the uncertainty. For the BBHs, the bulk of the simulations yields uncertainties in the range $\sim [0.5 - 10]$~\%. Owing to their shorter inspiral, heavy BBH have larger uncertainties, above 5\% at SNR 12 and above 8\% at SNR 10 or lower. 

The source-frame chirp mass is measured with uncertainties that are an order of magnitude larger for the BNSs, NSBHs and BBHs, and a factor of two larger for hBBHs, than those for the detector-frame chirp mass, due to large uncertainties on the luminosity distance measurements (thus redshift measurements). These uncertainties would decrease if an EM counterpart is identified, and provides a better redshift measurement.

The mass ratio can only be significantly constrained for sources with visible spin precession (NSBHs) or long inspirals (BNSs).
At SNR 12, the 90\% can be as small as $\sim 0.07$ for our NSBHs, and $[0.3 - 0.5]$ for BNS.

Finally, the effective inspiral spins can be constrained, obtaining posterior distributions a factor of many narrower than the prior for systems with large mass ratios and spin precession. For those, at SNR 10 the uncertainty can be as small as $\sim 0.1$, although $\sim 0.4$ is more common (the prior width is 0.89). Using the K-L divergence we quantified the amount of information provided by the data. For systems with at least one BH, that number is usually above 1 nat.
For BNS, we have used a more restrictive prior (the prior width is 0.05), and obtained that the data usually yield only some information. 
At SNR 12, the typical K-L divergence we have found for BNSs is 0.1 nats. As the SNR decreases, the data does not allow to significantly update the  prior and the K-L divergence can get values below 0.1 nats. 

Based on the uncertainties mentioned above, it might be occasionally possible to associate a marginal source to a specific astrophysical class (e.g. BBH rather than BNS). 
It is important to remember that even if each event individually is not particularly informative, the whole population of marginal events can be used and contribute to the astrophysical inference of the underlying population. This can be done even if one is not certain about their astrophysical origin, since the probability that the event is astrophysical can be folded in the analysis~\cite{1809.03815}.

The posterior samples produced for this study have been made available in a public repository~\cite{yiwen_huang_2018_1601151}.

\section{Acknowledgments}
Y.~H., K.~N. and S.~V.~acknowledge support of the MIT physics department through the Solomon Buchsbaum Research Fund, the National Science Foundation, and the LIGO Laboratory. 
H.~M. acknowledges support from the Science and Technology Facilities Council (STFC) and the Australian Research Council Center of Excellence for Gravitational Wave Discovery (OzGrav) through project number CE170100004, as well as travel support from the Royal Astronomical Society and the Institute of Physics. J.~V. is supported by STFC Grant No. ST/K005014/1.
LIGO was constructed by the California Institute of Technology and Massachusetts Institute of Technology with funding from the National Science Foundation and operates under cooperative agreement PHY-0757058. 
The authors would like to thank Hsin-Yu Chen, Thomas Dent, Carl-Johan Haster, Erik Katsavounidis, and Ryan Lynch for useful discussions, and Chad Hanna and Kipp Cannon for help with running the recoloring routines of \textit{gstlal}.
The authors  acknowledge usage of LIGO Data Grid clusters. 
This is LIGO Document Number P1800309.
\appendix
\section{Sampling convergence}\label{Sec.AppendixSampling}

As mentioned in the body of the paper, sampling algorithms are only guaranteed to return the correct posteriors in the limit in which they run for an infinite amount of time.
Especially at low SNRs, there is the risk that the sampler does not reach the right part of the parameter space, or gets stuck in a local and unrelated maxima.
We use the maximum value of the recovered log likelihood as a probe of the convergence of the run, by comparing it with the log likelihood corresponding to the exact waveform parameters.

For a well converged run, the difference:
$$\Delta\log{L}\equiv \log{L}_{\mathrm{injected}} - \mathrm{max}\left(\log{L}_{\mathrm{recovered}}\right)$$
should be slightly negative (and not exactly zero since the component of the noise that happens to be correlated across the network can contribute to the maximum recovered log likelihood.
Conversely, positive values would suggest that the code failed to collect all the evidence that was available, which would be indicative of a problem with convergence. 

\begin{figure}[htb]
  \centering
    \includegraphics[width=0.45\textwidth]{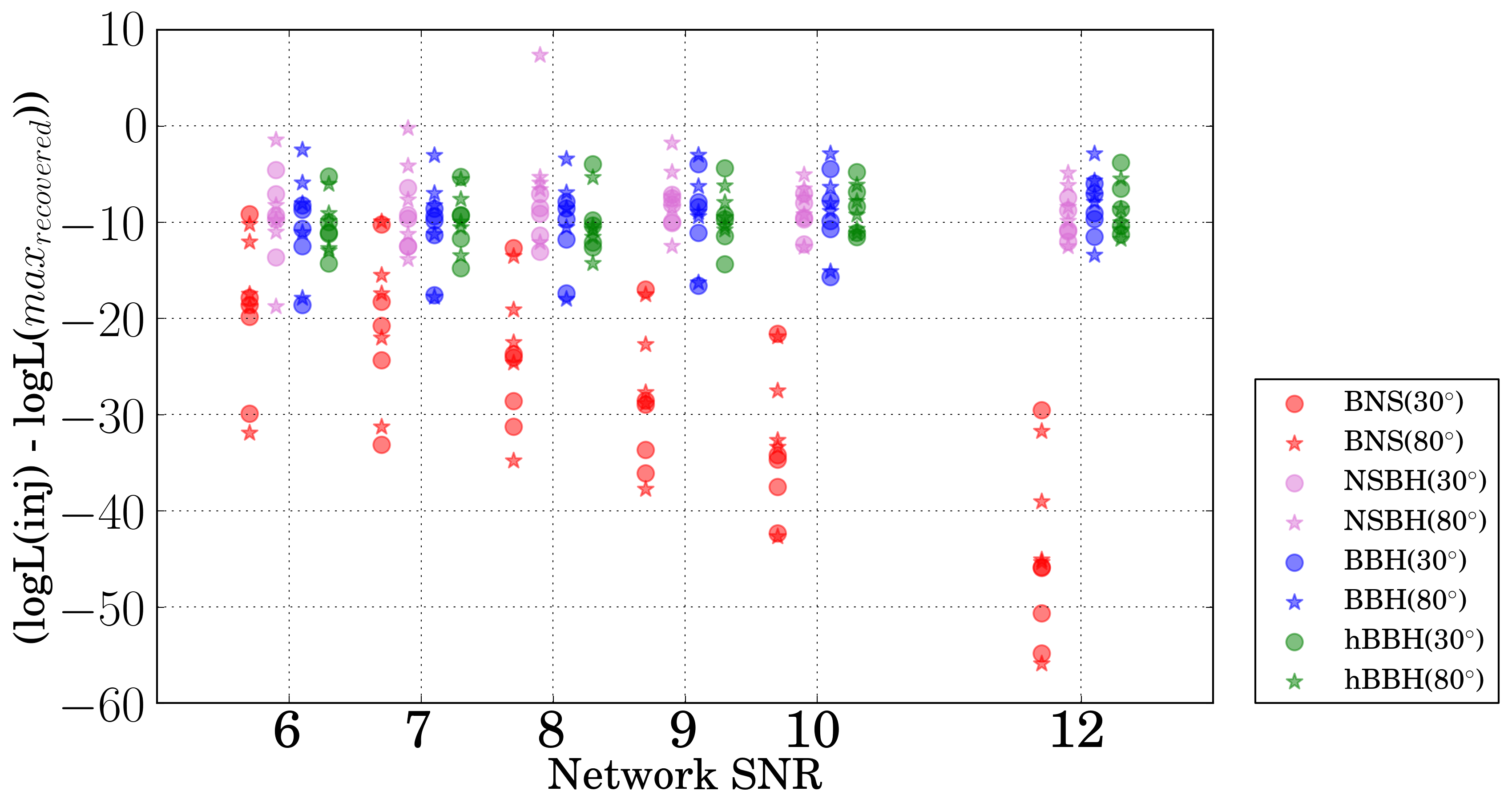}
    \caption{Log likelihood for injection minus the maximum recovered log likelihood vs. network SNR.}
  \label{Fig. logL_diff}
\end{figure}

We show $\Delta\log{L}$ for all sources in Fig.~\ref{Fig. logL_diff}. We find the expected behavior for all sources except for BNS, for which a downward trend is visible. We explain this difference with the fact that the current implementation of the ROQ likelihood in the \textit{lscsoft} algorithm repository has a known issue for long signals~\cite{Ng:2018neg}. Imperfections in the waveform reconstruction used in the ROQ basis can explain the different behavior of the BNS points.

To verify that the uncertainties we obtained for the BNS runs are meaningful, we have checked that the BNS runs do recover the correct SNR. Furthermore, we have run a small numbers of BNS sources (the ones used in Ref.~\cite{Ng:2018neg}) both with and without the ROQ likelihood and found that the uncertainties are similar. 

\section{Arrival Time}\label{Sec.AppendixTime}

\begin{figure}[!htbp]
  \centering
    \includegraphics[width=0.45\textwidth]{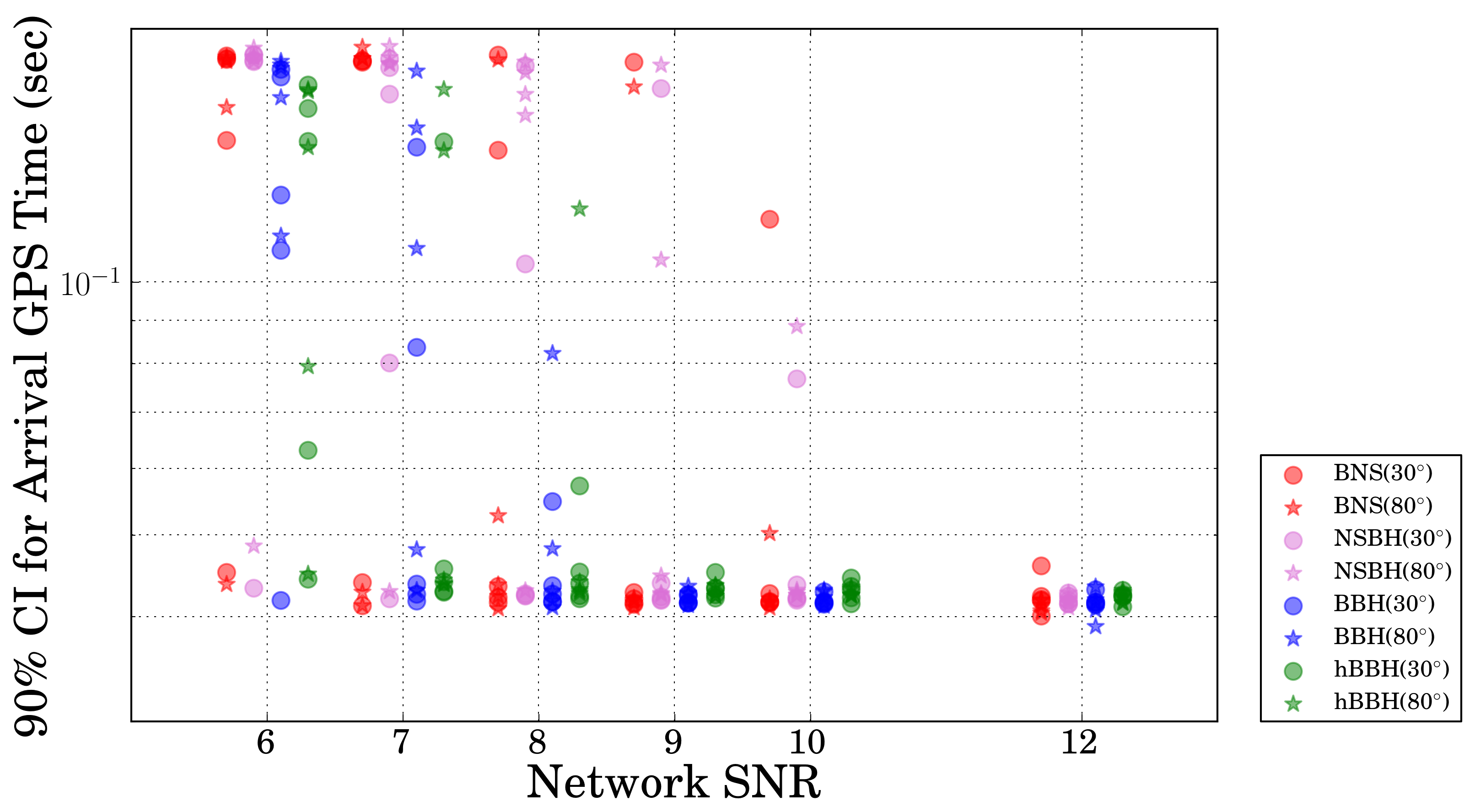}
    \caption{Relative 90\% credible interval of the marginalized posteriors of the arrival GPS time vs. network SNR. }
  \label{Fig. arrival time}
\end{figure}

Here we report the 90\% CI for the arrival GPS time, Fig. ~\ref{Fig. arrival time}. Time coincidence may be helpful in making an association of the GW signal with an EM counterpart. We find that the arrival time can be determined with uncertainties below 0.05 seconds for all sources at SNR of 12, and even at SNRs as low as 6, for some sources. 

\section{K-L Divergence}\label{Sec.K-L Divergence}

Here we report the median of K-L divergence (in nats) over 5 GPS times  for luminosity distance $d_L$, arrival GPS time, right ascension $\alpha$, declination $\delta$, detector-frame chirp mass $\mathcal{M}$, source-frame chirp mass \mcsource, mass ratio q, effective spin \chieff, and effective precessing spin \chip, for the four morphologies and the two inclinations, \Crefrange{Table.KLBNS}{Table.KLhBBH}.

\begin{table}[!htbp]
\centering
\begin{tabularx}{0.49\textwidth}{WNNNNNN}
\toprule\toprule
 Network SNR & 6 & 7 & 8 & 9 & 10 & 12 \\
\midrule
$d_L$ (inc=$30^\circ$) & 0.01 & 0.06 & 4.43 & 5.28 & 5.83 & 6.67 \\
$d_L$ (inc=$80^\circ$) & 0.01 & 0.02 & 3.86 & 5.17 & 5.72 & 6.68 \\
time (inc=$30^\circ$) & 0.01 & 0.03 & 2.61 & 2.78 & 2.82 & 2.85 \\
time (inc=$80^\circ$) & 0.00 & 0.02 & 2.13 & 2.83 & 2.86 & 2.92 \\
$\alpha$ (inc=$30^\circ$) & 0.01 & 0.02 & 0.39 & 0.46 & 0.45 & 0.90 \\
$\alpha$ (inc=$80^\circ$) & 0.00 & 0.01 & 0.41 & 0.41 & 0.53 & 0.76 \\
$\delta$ (inc=$30^\circ$) & 0.01 & 0.02 & 0.28 & 0.59 & 0.53 & 0.98 \\
$\delta$ (inc=$80^\circ$) & 0.00 & 0.01 & 0.46 & 0.35 & 0.36 & 0.54 \\
$\mathcal{M}$ (inc=$30^\circ$) & 0.02 & 0.04 & 5.61 & 10.45 & 10.46 & 11.21 \\
$\mathcal{M}$ (inc=$80^\circ$) & 0.01 & 0.03 & 4.54 & 10.68 & 10.82 & 10.98 \\
\mcsource (inc=$30^\circ$) & 0.01 & 0.02 & 3.94 & 4.43 & 4.65 & 4.97 \\
\mcsource (inc=$80^\circ$) & 0.01 & 0.03 & 3.37 & 4.40 & 4.60 & 4.91 \\
q (inc=$30^\circ$) & 0.00 & 0.01 & 1.70 & 1.79 & 1.81 & 1.85 \\
q (inc=$80^\circ$) & 0.00 & 0.00 & 1.48 & 1.83 & 1.89 & 1.93 \\
\chieff (inc=$30^\circ$) & 0.01 & 0.01 & 0.05 & 0.08 & 0.07 & 0.09 \\
\chieff (inc=$80^\circ$) & 0.01 & 0.01 & 0.06 & 0.08 & 0.08 & 0.12 \\
\chip (inc=$30^\circ$) & 0.01 & 0.00 & 0.04 & 0.06 & 0.06 & 0.06 \\
\chip (inc=$80^\circ$) & 0.00 & 0.01 & 0.06 & 0.06 & 0.06 & 0.06 \\
\bottomrule\bottomrule
\end{tabularx}
\caption{Median of K-L divergence (in nats) over 5 GPS times for BNS. }
\label{Table.KLBNS}
\end{table}

 \begin{table}[!htbp]
\centering
\begin{tabularx}{0.49\textwidth}{WNNNNNN}
\toprule\toprule
 Network SNR & 6 & 7 & 8 & 9 & 10 & 12 \\
\midrule
$d_L$ (inc=$30^\circ$) & 0.01 & 0.67 & 4.23 & 5.20 & 5.83 & 6.65 \\
$d_L$ (inc=$80^\circ$) & 0.01 & 0.02 & 0.80 & 5.98 & 7.28 & 8.77 \\
time (inc=$30^\circ$) & 0.00 & 0.27 & 2.57 & 2.88 & 2.88 & 2.95 \\
time (inc=$80^\circ$) & 0.00 & 0.01 & 0.22 & 2.70 & 2.81 & 3.37 \\
$\alpha$ (inc=$30^\circ$) & 0.00 & 0.14 & 0.41 & 0.67 & 0.75 & 0.90 \\
$\alpha$ (inc=$80^\circ$) & 0.00 & 0.00 & 0.04 & 0.68 & 0.70 & 1.45 \\
$\delta$ (inc=$30^\circ$) & 0.00 & 0.21 & 0.48 & 0.68 & 0.73 & 0.92 \\
$\delta$ (inc=$80^\circ$) & 0.00 & 0.01 & 0.07 & 0.70 & 0.68 & 1.62 \\
$\mathcal{M}$ (inc=$30^\circ$) & 0.01 & 0.53 & 4.99 & 7.35 & 7.39 & 7.49 \\
$\mathcal{M}$ (inc=$80^\circ$) & 0.01 & 0.04 & 0.58 & 5.71 & 7.52 & 8.05 \\
\mcsource (inc=$30^\circ$) & 0.01 & 0.29 & 2.21 & 2.61 & 2.82 & 3.08 \\
\mcsource (inc=$80^\circ$) & 0.02 & 0.03 & 0.29 & 2.79 & 3.48 & 4.13 \\
q (inc=$30^\circ$) & 0.01 & 0.06 & 0.13 & 0.08 & 0.39 & 0.30 \\
q (inc=$80^\circ$) & 0.01 & 0.00 & 0.03 & 0.55 & 1.24 & 1.74 \\
\chieff (inc=$30^\circ$) & 0.01 & 0.29 & 2.30 & 2.85 & 2.89 & 3.20 \\
\chieff (inc=$80^\circ$) & 0.01 & 0.01 & 0.07 & 3.52 & 4.71 & 5.03 \\
\chip (inc=$30^\circ$) & 0.00 & 0.01 & 0.18 & 0.20 & 0.19 & 0.34 \\
\chip (inc=$80^\circ$) & 0.00 & 0.00 & 0.00 & 1.32 & 1.57 & 1.76 \\
\bottomrule\bottomrule
\end{tabularx}
\caption{Median of K-L divergence (in nats) over 5 GPS times for NSBH. }
\label{Table.KLNSBH}
\end{table} 

 \begin{table}[!htbp]
\centering
\begin{tabularx}{0.49\textwidth}{WNNNNNN}
\toprule\toprule
 Network SNR & 6 & 7 & 8 & 9 & 10 & 12 \\
\midrule
$d_L$ (inc=$30^\circ$) & 1.92 & 4.69 & 5.11 & 5.60 & 6.15 & 6.89 \\
$d_L$ (inc=$80^\circ$) & 0.02 & 2.09 & 5.43 & 5.71 & 6.60 & 7.76 \\
time (inc=$30^\circ$) & 0.95 & 2.19 & 2.53 & 2.94 & 3.11 & 3.02 \\
time (inc=$80^\circ$) & 0.01 & 1.05 & 2.60 & 2.90 & 2.94 & 2.97 \\
$\alpha$ (inc=$30^\circ$) & 0.08 & 0.18 & 0.38 & 0.70 & 0.71 & 0.89 \\
$\alpha$ (inc=$80^\circ$) & 0.00 & 0.14 & 0.38 & 0.62 & 0.55 & 0.61 \\
$\delta$ (inc=$30^\circ$) & 0.15 & 0.45 & 0.41 & 0.52 & 0.59 & 0.68 \\
$\delta$ (inc=$80^\circ$) & 0.00 & 0.15 & 0.56 & 0.67 & 0.69 & 0.82 \\
$\mathcal{M}$ (inc=$30^\circ$) & 1.55 & 3.28 & 4.18 & 4.94 & 5.48 & 5.58 \\
$\mathcal{M}$ (inc=$80^\circ$) & 0.03 & 1.24 & 4.26 & 4.76 & 5.09 & 5.44 \\
\mcsource (inc=$30^\circ$) & 0.78 & 2.21 & 2.46 & 3.09 & 3.64 & 4.27 \\
\mcsource (inc=$80^\circ$) & 0.01 & 0.90 & 2.72 & 3.26 & 4.06 & 5.02 \\
q (inc=$30^\circ$) & 0.13 & 0.27 & 0.47 & 0.57 & 0.60 & 0.63 \\
 (inc=$80^\circ$) & 0.02 & 0.14 & 0.24 & 0.51 & 0.45 & 0.66 \\
\chieff (inc=$30^\circ$) & 0.12 & 0.27 & 1.25 & 1.44 & 1.65 & 1.79 \\
\chieff (inc=$80^\circ$) & 0.01 & 0.26 & 1.01 & 1.52 & 1.78 & 2.05 \\
\chip (inc=$30^\circ$) & 0.01 & 0.07 & 0.06 & 0.08 & 0.10 & 0.09 \\
\chip (inc=$80^\circ$) & 0.00 & 0.01 & 0.25 & 0.75 & 0.79 & 1.38 \\
\bottomrule\bottomrule
\end{tabularx}
\caption{Median of K-L divergence (in nats) over 5 GPS times for BBH. }
\label{Table.KLBBH}
\end{table} 

 \begin{table}[!htbp]
\centering
\begin{tabularx}{0.49\textwidth}{WNNNNNN}
\toprule\toprule
 Network SNR & 6 & 7 & 8 & 9 & 10 & 12 \\
\midrule
$d_L$ (inc=$30^\circ$) & 1.25 & 4.40 & 5.06 & 5.66 & 6.21 & 6.94 \\
$d_L$ (inc=$80^\circ$) & 1.18 & 3.87 & 4.84 & 5.41 & 6.06 & 7.22 \\
time (inc=$30^\circ$) & 0.49 & 2.62 & 2.69 & 2.93 & 2.88 & 3.43 \\
time (inc=$80^\circ$) & 0.55 & 2.20 & 2.63 & 2.97 & 2.86 & 3.17 \\
$\alpha$ (inc=$30^\circ$) & 0.11 & 0.29 & 0.39 & 0.57 & 0.74 & 1.25 \\
$\alpha$ (inc=$80^\circ$) & 0.09 & 0.25 & 0.40 & 0.60 & 0.47 & 1.02 \\
$\delta$ (inc=$30^\circ$) & 0.04 & 0.40 & 0.49 & 0.75 & 0.98 & 1.44 \\
$\delta$ (inc=$80^\circ$) & 0.04 & 0.26 & 0.30 & 0.50 & 0.52 & 1.19 \\
$\mathcal{M}$ (inc=$30^\circ$) & 0.53 & 1.63 & 1.74 & 2.90 & 3.25 & 3.35 \\
$\mathcal{M}$ (inc=$80^\circ$) & 0.57 & 1.45 & 2.11 & 2.66 & 2.79 & 3.40 \\
\mcsource (inc=$30^\circ$) & 0.18 & 1.16 & 1.47 & 1.94 & 2.26 & 2.91 \\
\mcsource (inc=$80^\circ$) & 0.20 & 0.85 & 1.29 & 1.72 & 2.14 & 3.05 \\
q (inc=$30^\circ$) & 0.13 & 0.54 & 0.60 & 0.67 & 0.79 & 0.91 \\
q (inc=$80^\circ$) & 0.08 & 0.21 & 0.41 & 0.58 & 0.66 & 0.81 \\
\chieff (inc=$30^\circ$) & 0.33 & 1.43 & 1.60 & 1.52 & 1.46 & 1.75 \\
\chieff (inc=$80^\circ$) & 0.27 & 0.72 & 1.10 & 1.47 & 1.51 & 1.71 \\
\chip (inc=$30^\circ$) & 0.02 & 0.08 & 0.10 & 0.10 & 0.10 & 0.11 \\
\chip (inc=$80^\circ$) & 0.01 & 0.03 & 0.09 & 0.19 & 0.21 & 0.49 \\
\bottomrule\bottomrule
\end{tabularx}
\caption{Median of K-L divergence (in nats) over 5 GPS times for hBBH. }
\label{Table.KLhBBH}
\end{table} 

Numbers close to zero imply the data is not informative about that parameter (at a given SNR). Conversely, large K-L divergence implies the prior and posteriors are significantly different.

\chip is notoriously hard to measure, even for louder events~\cite{GW150914-PARAMESTIM,2016arXiv160601210T,GW151226-DETECTION,2017PhRvL.119n1101A,2017PhRvL.118v1101A,2017ApJ...851L..35A,2017PhRvL.119y1103V,O1-BBH,LIGOScientific:2018jsj}. The tables below show that for weak events we nearly always recover the prior, which is why we have not reported  \chip uncertainties in the body of the paper.

\clearpage
\bibliographystyle{apsrev4-1}
\bibliography{SalvosBib}

\end{document}